\documentclass[a4paper,11pt]{article}

\usepackage{geometry}										
\usepackage[english]{babel}									
\usepackage[utf8]{inputenc}									
\usepackage[T1]{fontenc}									

\usepackage{lmodern}										

\usepackage{amssymb,amsfonts,amsmath,amsthm}
\usepackage{booktabs}
\usepackage{hyphenat}
\usepackage{booktabs}
\usepackage{bbm}
\usepackage{multirow}
\usepackage{graphicx}
\usepackage{algpseudocode}
\usepackage{algorithm}

\makeatletter
\let\oldabs\abs
\def\abs{\@ifstar{\oldabs}{\oldabs*}}
\let\oldnorm\norm
\def\norm{\@ifstar{\oldnorm}{\oldnorm*}}
\makeatother

\newcommand{\Prob}[1]{{\mathbb P}\left( #1 \right)}	
\renewcommand{\vec}[1]{\boldsymbol{#1}}
\newcommand{\dt}{{\delta t}}
\newcommand{\yhat}{{\hat y}}

\usepackage{authblk}										

\title{Emulating the dynamics of complex systems using autoregressive models on manifolds (mNARX)}
\author[1]{Styfen Schär \thanks{styfen.schaer@ibk.baug.ethz.ch}}
\author[1]{Stefano Marelli\thanks{marelli@ibk.baug.ethz.ch}}
\author[1]{Bruno Sudret\thanks{sudret@ethz.ch}}
\affil[1]{Chair of Risk, Safety and Uncertainty Quantification, ETH Z\"{u}rich, Switzerland}

\date{\today}

\usepackage{microtype}										

\usepackage{indentfirst}									

\usepackage{caption}
\usepackage{subcaption}

\usepackage{natbib}

\usepackage{hyperref}
\hypersetup{
pdftitle = {Emulating the dynamics of complex systems using autoregressive models on manifolds (mNARX)},
pdfauthor = {Styfen Sch\"ar, Stefano Marelli, Bruno Sudret},
pdfsubject = {},
pdfkeywords = {Wind turbine simulation, autoregressive models, manifold learning, mNARX},
colorlinks = false,                                                                                                                                                                          
}

\usepackage[capitalize]{cleveref}								
\creflabelformat{equation}{#2(#1)#3}								
\crefrangelabelformat{equation}{#3(#1)#4 to #5(#2)#6}						
\labelcrefformat{subequation}{#2(#1)#3}								
\labelcrefrangeformat{subequation}{#3(#1)#4 to #5(#2)#6}					

\graphicspath{{./figures}} 

\begin{document}

\maketitle

\begin{abstract}
We propose a novel surrogate modelling approach to efficiently and accurately approximate the response of complex dynamical systems driven by time-varying exogenous excitations over extended time periods. 
Our approach, namely \emph{manifold nonlinear autoregressive modelling with exogenous input} (mNARX), involves constructing a problem-specific exogenous input manifold that is optimal for constructing autoregressive surrogates. 
The manifold, which forms the core of mNARX, is constructed incrementally by incorporating the physics of the system, as well as prior expert- and domain- knowledge. 
Because mNARX decomposes the full problem into a series of smaller sub-problems, each with a lower complexity than the original, it scales well with the complexity of the problem, both in terms of training and evaluation costs of the final surrogate. 
Furthermore, mNARX synergizes well with traditional dimensionality reduction techniques, making it highly suitable for modelling dynamical systems with high-dimensional exogenous inputs, a class of problems that is typically challenging to solve.

Since domain knowledge is particularly abundant in physical systems, such as those found in civil and mechanical engineering, mNARX is well suited for these applications. 
We demonstrate that mNARX outperforms traditional autoregressive surrogates in predicting the response of a classical coupled spring-mass system excited by a one-dimensional random excitation. 
Additionally, we show that mNARX is well suited for emulating very high-dimensional time- and state-dependent systems, even when affected by active controllers, by surrogating the dynamics of a realistic aero-servo-elastic onshore wind turbine simulator.

In general, our results demonstrate that mNARX offers promising prospects for modelling complex dynamical systems, in terms of accuracy and efficiency.
\end{abstract}

\section{Introduction}\label{sec:introduction}
Modelling and predicting the behaviour of dynamical systems is a fundamental yet challenging task encountered in engineering and applied sciences.
The goals in modelling such systems are diverse and encompass for instance gaining deeper insight into the system dynamics to uncover its governing equations, e.g., for system control purposes \cite{levin_1996}.
For predictive maintenance \cite{langeron_2021}, digital twins \cite{edington_2023}, or damage detection \cite{mattson_2006}, the goal is to predict the future state of a system based on past observations and measurements. 
For the purpose of uncertainty quantification \cite{mai_2017, bhattacharyya_2020}, reliability analysis \cite{garg_2022}, or design optimization \cite{deshmukh_2017}, an attempt is made to reduce the cost of system response analysis.
This is achieved by replacing the system with an inexpensive-to-evaluate surrogate model.

Regardless of their specific goals, all of these examples share autoregressive modelling as a tool for describing the evolution in time of the system under study. 
This is due to the ability of autoregressive models to represent time-dependent data using past observations or predictions. 
A special class of autoregressive models is that of exogenous input autoregressive models (ARX), which incorporate exogenous inputs such as time-dependent loads or control signals to improve their predictive power.
ARX models are usually implemented as nonlinear ARX models (NARX), which consider nonlinear relationships between inputs and outputs for a more accurate representation of complex systems. For a comprehensive introduction to NARX models, the reader is referred to \cite{billings_2013}.

Many variants of NARX models have been developed and successfully applied in various fields.
For example, nonlinear polynomial models with exogenous inputs have been used to describe the evolution of magnetic activity due to solar winds \cite{balikhin_2001}, gas turbine shaft speeds \cite{chiras_2001}, or frictional dynamics \cite{wan_2008}.
\cite{mai_2016} and \cite{spiridonakos_2015} used them to create flexible surrogates for modelling dynamical systems under uncertain excitation. 
Neural network-based ARX models have been used with great success to reduce the measurement error of microelectromechanical systems \cite{li_2021}, in the context of fault detection in wind turbines \cite{schlechtingen_2011}, or for vibration control \cite{song_2022}.
Other popular algorithms used in the context of dynamical systems are state-space models \cite{zhang_2019}, or vector autoregressive models \cite{lutkepohl_2005}.
Classic algorithms from machine learning and surrogate modelling have also been reformulated to model dynamical systems, such as Gaussian process modelling \cite{murray_1999, koziel_2014, worden_2018, kocijan_2012} or support vector regression \cite{rankovic_2014}.

Despite the relatively rich literature on the topic, emulating the response of dynamical systems over extended time periods can still pose a major challenge. 
The difficulty often comes from the complexity of real-world systems, which can involve nonlinear springs, dampers, coupling, and controllers. 
These factors can result in a system response that is nonlinear, non-differentiable, or even discontinuous with respect to the exogenous input, making the approximation of the response challenging \cite{kerschen_2006}.
In these cases, classical NARX algorithms often fail, or require a significant amount of data to make accurate predictions. 
In most cases, this is because these algorithms rely on some form of regularity in the input-output state-space mapping, such as smoothness, symmetry, or stationarity.

In practice, the regularity assumptions may not hold when the system dynamics are complex, rendering traditional modelling methods ineffective. 
This challenge can be addressed by working in a different space in which these assumptions are satisfied. 
For example, Calandra et al. \cite{calandra_2016} demonstrated the effectiveness of a classical surrogate model even in the case of discontinuous functions, when a suitable feature space with smooth input-output mappings is constructed. However, identifying such a space can be difficult.

To solve this problem, various methods have been proposed to identify nonlinear transforms that map the input data into spaces better suited for surrogate modelling.
Recently, autoencoders have emerged as a powerful tool for this purpose. For instance, Champion et al. \cite{champion_2019} used autoencoders to identify a reduced set of coordinates that simplifies the identification of system dynamics, while Lee and Carlberg employed deep convolutional autoencoders to project dynamical systems onto nonlinear manifolds \cite{lee_2019}.
Autoencoders were also used by Simpson et al. \cite{simpson_2021} to construct reduced-order models of nonlinear dynamical systems.

A comprehensive review of other popular methods for identifying linear or nonlinear subspaces in the context of surrogate modelling can be found in \cite{lataniotis_2020}.
However, while lower-dimensional spaces are often preferred for surrogate modelling due to the curse of dimensionality \cite{hutchison_2005}, they can sometimes result in unfavorable topologies for surrogate modelling \cite{lataniotis_2020}. 
In summary, identifying an input manifold that facilitates state-space surrogate modelling is known to significantly improve the predictive capabilities of classical emulation techniques. 

With this in mind, we propose a novel surrogate modelling technique, called \emph{manifold nonlinear autoregressive with exogenous input} (mNARX) modelling. 
This approach combines NARX modelling with a supervised incremental construction of a nonlinear exogenous input manifold. 
mNARX can accurately approximate complex dynamical systems, even when their response is dependent on high-dimensional exogenous inputs. 
Moreover, mNARX achieves this with high computational efficiency and stability over extended time periods.

The organization of the paper is as follows: In Section~\ref{sec:mNARX}, we describe the general rationale of mNARX, including its individual components and the details of the algorithm.
In Section~\ref{sec:applications}, we address the construction of mNARX from an existent training dataset (the \textit{experimental design}).
To validate its performance, we present two case studies: a coupled spring-mass system with a one-dimensional exogenous input, and a numerical aero-servo-elastic wind turbine simulator with a control system and high-dimensional exogenous input (three-dimensional wind field sampled over 10~minutes).
Finally, in Section~\ref{sec:discussion_and_conclusion}, we provide concluding remarks on the mNARX algorithm and discuss potential extensions to enhance its performance and applicability.

\section{Manifold autoregressive with exogenous input modelling}\label{sec:mNARX}
Our goal is to introduce a novel algorithm, namely mNARX (for manifold nonlinear autoregressive with exogenous input) modelling, which can emulate the response of deterministic systems, excited by time-varying exogenous inputs. 
We describe such systems mathematically as
\begin{equation}\label{eq:original_mapping_init_cond}
	y(t) = \mathcal{M}(\vec{x}(\mathcal{T} \le t), \vec{\beta}),
\end{equation}
where $y(t)\in \mathbb{R}$ is one scalar component of the response of the system at time $t \in \mathcal{T}$, $\vec{x}(t)\in \mathbb{R}^M$ is the $M$-dimensional vector of exogenous inputs and $\vec{\beta}$ describes the initial conditions of the system.
For notational simplicity, we hereafter omit $\vec{\beta}$ in our notation unless strictly necessary, and use the notation:
\begin{equation}\label{eq:original_mapping}
	y(t) = \mathcal{M}(\vec{x}(\mathcal{T} \le t)).
\end{equation}

Although the system response can also be multivariate, for notation simplicity, we assume a univariate response.
Further, we focus on discrete-time problems, which means that each element of the time axis $\mathcal{T}=\{0, \dt, 2\dt, \dots, (N-1)\dt\}$ is an integer multiple of the time increment $\delta t$. 
However, the algorithm presented can also be applied to continuous-time problems.
The notation $\bullet(\mathcal{T} \le t)$ indicates that the system is causal, meaning the response depends on past and current inputs, but not on future inputs.

Our objective with the mNARX algorithm is to construct a surrogate on a finite set of system inputs $\vec{x}^{(i)}$ and corresponding system outputs $\vec{y}^{(i)}$ called the \emph{experimental design} (ED):
\begin{equation}\label{eq:experimental_design}
	\mathcal{D} = \left\{ \left( \vec{x}^{(i)}, \vec{y}^{(i)} \right) , \vec{x}^{(i)} \in \mathbb{R}^{N \times M},  \vec{y}^{(i)} = \mathcal{M}(\vec{x}^{(i)}) \in \mathbb{R}^{N}, i=1, \dots, N_\text{ED} \right\}.
\end{equation}
The surrogate $\hat{\mathcal{M}}$ shall emulate the system response (hereinafter the \textit{model prediction}) over a long time period, based solely on the exogenous inputs:
\begin{equation}\label{eq:approx_mappinsg}
	y(t) = \mathcal{M}(\vec{x}(\mathcal{T} \le t), \vec{\beta}) \approx \hat{\mathcal{M}}(\vec{x}(\mathcal{T} \le t), \hat{\vec{\beta}}).
\end{equation}

This can be challenging when the mapping from the exogenous input to the system output is highly nonlinear, the system has a long memory, the dimensionality of $\vec{x}(t)$ is high, or when the ED is small. To address these issues, mNARX uses a multi-step surrogate modelling approach based on two key components. 
The first is NARX modelling, introduced in Section~\ref{sec:ar_modelling}. 
The second is the incremental construction of an exogenous input space, namely the exogenous input manifold, that can encode prior information on the physics of the system under investigation, and that is more suitable for autoregressive modelling. A detailed description of this process is described in Section~\ref{sec:manifold_construction}. 
The full mNARX algorithm is summarized in Section \ref{sec:algorithm}.

\subsection{Autoregressive with exogenous input modelling}\label{sec:ar_modelling}
We employ AutoregRessive with eXogenous input (ARX) models as the basic tool to approximate the response of dynamical systems. 
At each time instant $t$ on a discretized time axis $\widetilde{\mathcal{T}}=\{t_0, t_0+\dt, \dots, (N-1)\dt\}$, the ARX model $\tilde{\mathcal{M}}^\text{}$ maps an input vector $\vec{\varphi}(t) \in \mathbb{R}^{M_\varphi}$ to its corresponding model output $y(t) \in \mathbb{R}$, which we express as
\begin{equation}\label{eq:NARX}
	y(t) = \tilde{\mathcal{M}}^\text{}(\vec{\varphi}(t), \mathbf{c}).
\end{equation}
The input vector $\vec{\varphi}(t)$ contains information on the system state variables at different time steps, and $\tilde{\mathcal{M}}^\text{}$ is a parametric model characterized by the finite set of model parameters $\mathbf{c}$. 
Typical examples of parametric ARX models include e.g. polynomials and Gaussian processes, or any mathematical function that is relatively simple and fast to evaluate. Note that when the model is nonlinear, it is often referred to as a Nonlinear ARX model (NARX).

Once the specific structure of the autoregressive model in Eq.~\eqref{eq:NARX} is chosen, the optimal set of parameters $\mathbf{c}$ needs to be calibrated from a dataset representative of the specific problem at hand, a process detailed in Section~\ref{sec:ar_modelling_fitting}.

\subsubsection{Calibration of an ARX model}\label{sec:ar_modelling_fitting}
To calibrate the coefficients $\mathbf{c}$ in Eq.~\eqref{eq:NARX}, we first define the so-called \textit{design matrix}:
\begin{equation}\label{eq:regression matrix}
	\vec{\Phi} = \begin{pmatrix}
		\vec{\varphi}(t_0) \\
		\vec{\varphi}(t_0+\delta t) \\
		\vdots \\
		\vec{\varphi}((N-1)\delta t)
	\end{pmatrix},
\end{equation}
which is constructed by stacking the input vectors of all time steps.     
It is formed from the time-dependent exogenous input $\vec{x}(t) \in \mathbb{R}^{M}$ and the output of the model itself $y(t) \in \mathbb{R}$. 

In its extended form, every single row of the design matrix reads
\begin{equation}\label{eq:exo_input}
	\begin{split}
		\vec{\varphi}(t) = \{
		&y(t - \ell_{1}^{y}), y(t - \ell_{2}^{y}), \dots, y(t - \ell_{n_y}^{y}), \\
		&x_1(t - \ell_{1}^{x_1}), x_1(t - \ell_{2}^{x_1}), \dots, x_1(t - \ell_{n_{x_1}}^{x_1}), \\ 
		&x_2(t - \ell_{1}^{x_2}), x_2(t - \ell_{2}^{x_2}), \dots, x_2(t - \ell_{n_{x_2}}^{x_2}),    \\
		&\dots, \\
		&x_{M}(t - \ell_{1}^{x_{M}}), x_{M}(t - \ell_{2}^{x_{M}}), \dots, x_{M}(t - \ell_{n_{x_{M}}}^{x_{M}})\},
	\end{split}
\end{equation}
where $\ell_{i}^{y} \in \{ \dt, 2\dt, \dots, (N-1)\dt \}$ are called autoregressive lags, and $\ell_{i}^{x_j} \in \{0,  \dt, 2\dt, \dots, (N-1)\dt \}$ are the exogenous input lags. 
We use the term \emph{lag} to refer to the time delay between the output at the current time and the value of a variable included in the model.
The minimum possible autoregressive lag is strictly larger than zero to preserve causality ($\ell_{i}^{y} > 0$). 
On the other hand, no such limitation is needed for the exogenous input, which means that an immediate effect of the exogenous input on the system response can be modelled ($\ell_{i}^{x_j} \ge 0$). 
It is also important to note that both the autoregressive and exogenous input lags can be non-contiguous, e.g., $\ell_{1}^{y} + \dt \ne \ell_{2}^{y}$, which means that it is not necessary to include the full set of possible lags in the model. 
Note that, since we use lagged time steps of the exogenous input and the autoregressive input, the number of time steps in $\vec{\varphi}$ is less than or equal (if no lags are included) to the number of time steps in $\vec{x}$ ($\widetilde{\mathcal{T}} \subseteq \mathcal{T}$).

There is no strict limit on the maximum lag (except the number of time steps available), but in practice, it is restricted to a reasonable size to keep the cardinality of $\vec{\varphi}(t)$, and hence typically that of $\vec{c}$ in Eq.~\eqref{eq:NARX} manageable. 
Because a large number of lags may be required to surrogate a complex system, in Section~\ref{sec:manifold_construction} we propose to substitute a large and complex surrogate by a chain of multiple simpler ones and thus reduce the number of lags required. 
Since the cardinality of $\vec{\varphi}(t)$ also increases with the number of exogenous inputs $M$, we discuss in Section~\ref{sec:dimensionality_reduction} how one can exploit standard dimensionality reduction techniques to address this issue.

Due to the wide range of applications of NARX modelling, a heuristic trial-and-error process is often necessary to determine a suitable set of lags in Eq.~\eqref{eq:exo_input}. 
While in principle a systematic investigation capitalizing on techniques such as autocorrelation, cross-correlation, spectral analysis, grid search, or cross-validation could facilitate the identification of relevant lags, this topic lies outside the scope of this paper, and will be investigated in follow-up work.
In addition, using knowledge about the underlying system can serve as a good starting point for later refinements.

From Eq.~\eqref{eq:NARX} and \eqref{eq:exo_input} it becomes clear that we can map each input vector $\vec{\varphi}(t)$ to its corresponding model output $y(t)$ and therefore can formulate a time-dependent problem as an ordinary regression problem.
In this setting, these input-output pairs 
$\left\{\vec{\varphi}(t),y(t)\right\}$, herein \textit{samples}, 
no longer need to follow a temporal ordering.
This has two important implications for the ARX training process: first, samples from different simulations of the system, which we refer to as \emph{model realizations}, and therefore also the design matrices $\vec{\Phi}^{(i)}$ can be concatenated to form a larger and more informative design matrix:
\begin{equation}
	\vec{\Phi}^\text{ED} = \begin{pmatrix}
		\vec{\Phi}^{(1)} \\
		\vdots \\
		\vec{\Phi}^{(N_\text{ED})}
	\end{pmatrix}
\end{equation} 
where the superscript $(i)$ indicates the index of the realization within the full experimental design (ED) of size $N_\text{ED}$.
Analogously we define an output vector $\vec{y}^\text{ED}$, whose component $\vec{y}^{(i)}$ correspond to $\vec{\Phi}^{(i)}$, as
\begin{equation}\label{eq:exp_design}
	\vec{y}^\text{ED} = \begin{pmatrix}
		\vec{y}^{(1)} \\ 
		\vdots \\
		\vec{y}^{(N_\text{ED})} 
	\end{pmatrix}.
\end{equation} 
The regression task is then performed on $\vec{\Phi}^\text{ED}$ and $\vec{y}^\text{ED}$.

A second implication of the lack of chronological ordering in the rows of the design matrix, is that it allows one to subsample from $\vec{\Phi}^\text{ED}$ and $\vec{y}^\text{ED}$:

\begin{equation}
	\vec{\Phi}_S = \begin{pmatrix}
		\vec{\Phi}^\text{ED}_{r_1} \\
		\vec{\Phi}^\text{ED}_{r_2} \\
		\vdots \\
		\vec{\Phi}^\text{ED}_{r_k}
	\end{pmatrix}
	, \;
	\vec{y}_S = \begin{pmatrix}
		\vec{y}^\text{ED}_{r_1} \\
		\vec{y}^\text{ED}_{r_2} \\
		\vdots \\
		\vec{y}^\text{ED}_{r_k}
	\end{pmatrix},
\end{equation} 
where the $k$ indices $r_i \in \{1, 2, \dots, |\vec{y}^\text{ED}|\}$, with $|.|$ denoting the set cardinality, are randomly or deterministically drawn. Here, $\vec{\Phi}^\text{ED}_{i}$ refers to the $i^\text{th}$ row of $\vec{\Phi}^\text{ED}$ and $\vec{y}^\text{ED}_{i}$ refers to the $i^\text{th}$ element of $\vec{y}^\text{ED}$. The model fitting is then performed on $\vec{\Phi}_S$ and $\vec{y}_S$.

In this paper, subsampling is introduced as a means to facilitate the model fitting process when dealing with a large number of highly correlated samples within each realization, without significantly compromising the quality of the surrogate.
In particular, random subsampling is a popular method because it can provide a representative subset of the full sample set. Reducing the correlation among the samples can also improve the conditioning of the regression problem and thus enhance numerical stability.
Optimal sampling can play an important role in the construction of surrogate models and is a wide field of research \cite{giunta_2003, goel_2008, simpson_2001, hampton_2015, dos_santos_2008, chkifa_2015}. A detailed discussion of it is therefore beyond the scope of this paper.

\subsubsection{ARX model prediction}\label{sec:ar_modelling_prediction}
As outlined in Section~\ref{sec:ar_modelling_fitting}, ARX models make use of past output time steps to generate predictions for future time steps. 
While the actual output of the simulations in the experimental design is used when fitting the model coefficients, this is not possible when predicting on unseen input data. 
During the prediction phase, the ARX surrogate $\hat{\mathcal{M}}$ generates a new time step prediction by using its previous predictions $\hat y(\mathcal{T} < t)$:
\begin{equation}\label{eq:NARX_prediction}
	\hat y(t) = \hat{\mathcal{M}}^\text{}(\vec{x}(\mathcal{T} \le t), \hat y(\mathcal{T} <t))
\end{equation}

The prediction process must be initialized with the output at time steps $t \in \{0, \dots, (n-1)\delta t\}$, where $n = \max(\{\ell_1^{\hat y}, \dots, \ell_{n_y}^{\hat y}\})$,  and the surrogate predicts only at the remaining time steps $t \in \{n\delta t, \dots, (N-1)\delta t\}$.
We denote this as 
\begin{equation}\label{eq:NARX_prediction_with_init}
	\hat y(t) = \hat{\mathcal{M}}^\text{}(\vec{x}(\mathcal{T} \le t), \hat y(\mathcal{T} <t), \vec{\beta}),
\end{equation}
where $\vec{\beta}$ represents the initial conditions which are the output values at the $n$ first time steps.
This differs from simulators where the initial conditions typically consist only of values at time zero, such as initial displacements, velocities, or accelerations.

In practice, the starting conditions for the prediction can vary, but it is common to set the initial output time steps to zero, or to any physically meaningful initial conditions for the application in question.
Some systems can exhibit a high level of sensitivity to their initial conditions, and even a small change in these conditions can result in vastly different predictions of the system evolution. 
However, this is also the case for the corresponding full-physics-based model, which returns different output values if $\vec{\beta}$ changes.
To ensure robust model validation, particularly when comparing individual output traces of the surrogate and the simulator for a given input, it is recommended to initialize the surrogate with the true output. For predicting new data when the true output is not available, the surrogate should be initialized to any sensible value as mentioned earlier. It should be noted that this initialization consideration is not unique to the ARX methodology, but is ubiquitous in the modelling of dynamic systems.

\subsection{Constructing an exogenous input manifold}\label{sec:manifold_construction}
Constructing a surrogate in the form of Eq.~\eqref{eq:approx_mappinsg} for complex models $\tilde{\mathcal{M}}: \vec{x}(\mathcal{T} \le t) \to \vec{y}(t)$ can be a challenging task.
To address this, we introduce a feature space, or \emph{input manifold}, denoted by $\vec{\zeta} \in \mathbb{R}^{N \times M_\zeta}$, on which the mapping reads
\begin{equation}
	\label{eqn:manifold mapping}
	\tilde{\mathcal{M}}: \vec{\zeta}(\mathcal{T} \le t) \to y(t) .
\end{equation}
The exogenous input manifold $\vec{\zeta}(\mathcal{T} \le t)$ has two main goals: i) it has a manageable dimensionality, so that it can be handled efficiently with classical multi-dimensional NARX techniques described in Section~\ref{sec:ar_modelling} and ii) it capitalizes on prior knowledge about the system dynamics to provide a simpler mapping to the final quantity of interest $y(t)$. 
While the first goal is well known and can be achieved through classical dimensionality reduction techniques, such as frequency filtering or principal component analysis \cite{lataniotis_2020}, the second relies on decomposing the original problem in a sequence of sub-problems of increasing complexity.

In a more abstract sense, our objective is to find a function $\mathcal{F}: \vec{x} \to \vec{\zeta}$, that creates an input manifold more suitable for surrogate modelling than the original input space $\vec{x} \in \mathbb{R}^{N \times M}$. 
In general, the dimensionality of $\vec\zeta$ can be either smaller or larger than that of the original space $\vec{x}$ because it is known that smaller spaces may yield even more complex input-output mappings, as demonstrated in \cite{lataniotis_2020}. 
Instead, we aim at constructing a feature space that facilitates and simplifies the training of an accurate surrogate model.

In Section~\ref{sec:auxiliary_quantities}, we describe a supervised method for incrementally constructing such an input manifold $\vec{\zeta}$, by representation the relation $\vec{\zeta} = \mathcal{F}(\vec{x})$ using a composition of multiple simpler functions.
In Section~\ref{sec:dimensionality_reduction}, we provide further details on handling high-dimensional time-dependent inputs within the context of dynamical systems and discuss how this can be integrated into the manifold construction process.

\subsubsection{Auxiliary quantities}\label{sec:auxiliary_quantities}
Modelling the response of a dynamical system based solely on its raw exogenous inputs can be difficult, due to the presence of strong nonlinearities and coupling effects.
Such nonlinear or even discontinuous responses with respect to the raw input can be the result of control modules, nonlinear springs, dampers or coupling effects between system components.

mNARX addresses this issue by breaking down the modelling task into a series of intermediate steps, each with lower complexity than the full problem. 
Each modelling step corresponds to creating a new time-dependent feature, called an \emph{auxiliary quantity}, that provides valuable information about the system state.
These auxiliary quantities can be roughly classified into two categories: \emph{direct transformations} of the raw inputs, such as filtering or dimensionality reduction, and \emph{intermediate model responses}, which are calculated via a sequence of autoregressive models, such as control system responses, displacements at critical locations, etc.
The choice of the specific auxiliary quantities for a given system is based on prior expert knowledge about the system itself, and on the specific responses calculated by the modelling chain.
For instance, a domain expert may know that a certain transformation of the raw input or other auxiliary quantity is more informative than the original input time series.
Examples of auxiliary quantities can be moving averages of selected system outputs, integrals/derivatives, or even the response of an active control system.
In our experience, including exogenous variables such as control system states as additional exogenous inputs to the NARX model can significantly reduce its nonlinearity or eliminate discontinuities in the input-output mapping.

\begin{figure}
	\centering
	\includegraphics[width=10cm]{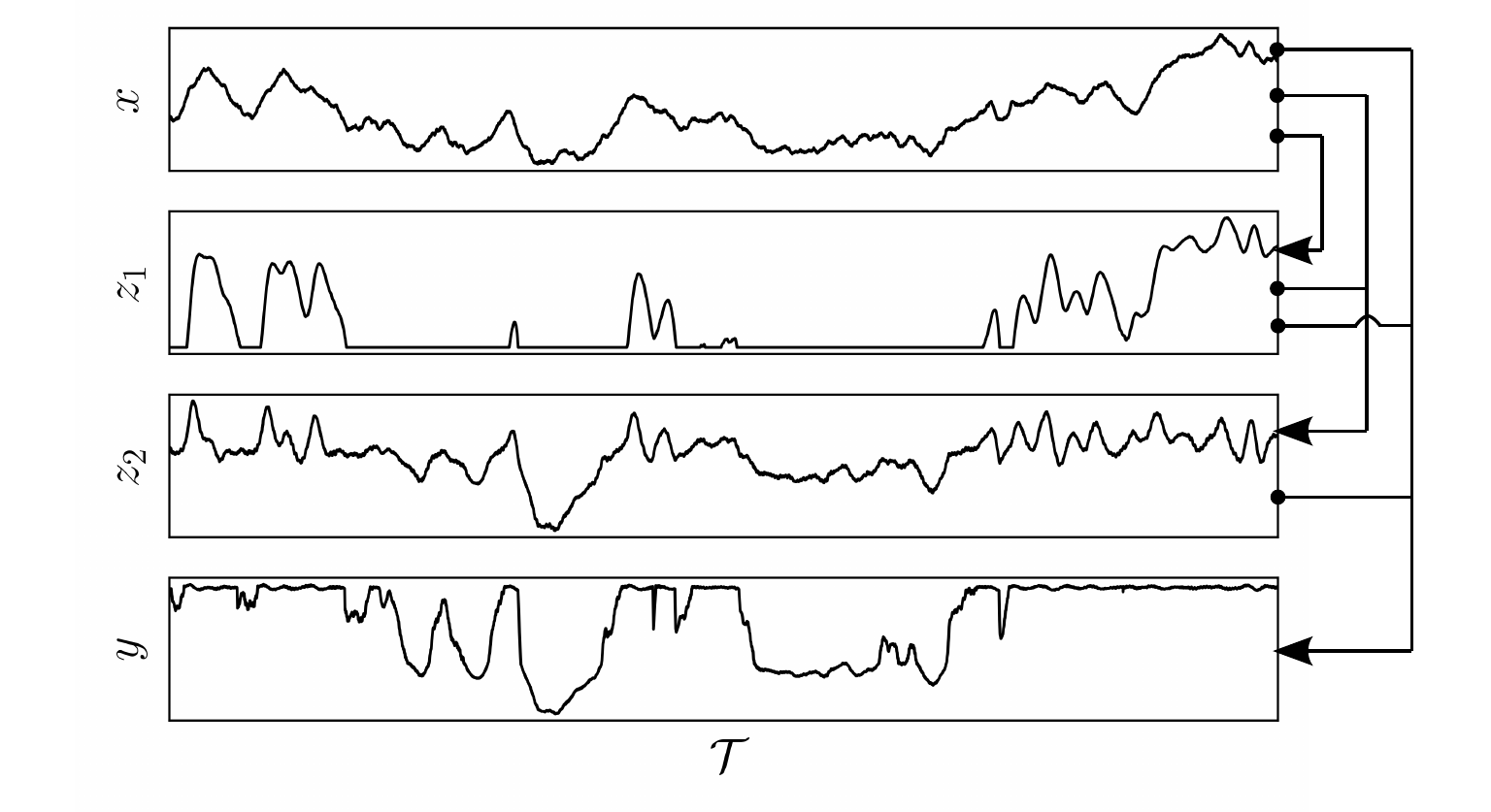}
	\caption{The figure shows the incremental construction of auxiliary quantities $z_i$ starting from the observed exogenous input $x$. With each iteration, the set of available features increases and the newly generated features provide more informative data, improving the mapping to the output $y$.}
	\label{fig:aux_quantities}
\end{figure}

From a formal perspective, each auxiliary quantity $\vec{z}_i(t) \in \mathbb{R}$, can be seen as a function $\mathcal{F}_i$ of all information available, including other auxiliary quantities $\vec{z}_{<i}(\mathcal{T} \le t)$, the raw exogenous input $\vec{x}(t) \in \mathbb{R}^M$ or past values of the auxiliary quantity $\vec{z}_{i}(\mathcal{T} <t)$ itself, as denoted in Eq.~\eqref{eq:aux_quantity} and illustrated in Fig.~\ref{fig:aux_quantities}:
\begin{equation}\label{eq:aux_quantity}
	\begin{split}
		\vec{z}_{1}(t) &= \mathcal{F}_1(\vec{x}(\mathcal{T} \le t), \vec{z}_{1}(\mathcal{T} <t)) \\
		\vec{z}_{2}(t) &= \mathcal{F}_2(\vec{z}_{1}(\mathcal{T} \le t), \vec{x}(\mathcal{T} \le t), \vec{z}_{2}(\mathcal{T} <t)) \\
		\vdots \\
		\vec{z}_{i}(t) &= \mathcal{F}_i(\vec{z}_{1}(\mathcal{T} \le t), \dots, \vec{z}_{i-1}(\mathcal{T} \le t), \vec{x}(\mathcal{T} \le t), \vec{z}_{i}(\mathcal{T} <t)).
	\end{split}
\end{equation}
We use the subscript to indicate the construction order of auxiliary quantities, where $\vec{z}_{i}$ represents the $i^{\text{th}}$ constructed quantity.
Note that each step in Eq.~\eqref{eq:aux_quantity} is closely related to Eq.~\eqref{eq:NARX_prediction} and that the auxiliary quantities can be viewed as an incremental extension of the exogenous input manifold.

In principle, auxiliary quantities can also be interdependent, so there is no clear order in which they have to be created. 
This scenario is shown in Eq.~\eqref{eq:aux_coupling}, where the $i^{\text{th}}$ feature is the result of the function $\mathcal{F}_i$ applied to the $j^{\text{th}}$ feature and the $j^{\text{th}}$ feature is the result of the function $\mathcal{F}_j$ applied to the $i^{\text{th}}$ one:
\begin{equation}\label{eq:aux_coupling}
	\begin{split}
		\vec{z}_{i}(t) &= \mathcal{F}_i(\vec{z}_{j}(\mathcal{T} \le t), \vec{x}(\mathcal{T} \le t), \vec{z}_{i}(\mathcal{T}<t)), \\
		\vec{z}_{j}(t) &= \mathcal{F}_j(\vec{z}_{i}(\mathcal{T} \le t), \vec{x}(\mathcal{T} \le t), \vec{z}_{j}(\mathcal{T}<t)).
	\end{split}
\end{equation}
In such a situation, we simply choose a leading quantity and alternately predict one time step at a time. 
For notational simplicity though, in this paper we will assume uncoupled auxiliary quantities, with a clear construction order.
Along the same lines, not all auxiliary quantities $\vec{z}_i(t)$ require all of the other quantities $\vec{z}_j(t), j<i$. 
Nevertheless, we will maintain the formulation in Eq.~\eqref{eq:aux_quantity} and Eq.~\eqref{eq:NARX_prediction} 
in its general form for consistency.

The incrementally growing set of exogenous inputs $\vec{\zeta}_i =\{ \vec{z}_{1}(\mathcal{T}\le t), \dots, \vec{z}_{i-1}(\mathcal{T} \le t), \vec{x}(\mathcal{T} \le t) \}$ is what we refer to as the exogenous input manifold.
Each auxiliary quantity can be thought of as a time-dependent system itself that depends on this manifold
\begin{equation}
	\vec{z}_i(t) = \mathcal{F}_i(\vec{\zeta}_i, \vec{z}_{i}(\mathcal{T} \le t)),
\end{equation}
which makes apparent that the function $\mathcal{F}_i$ can be regarded as an autoregressive model.
This stepwise enrichment of the manifold allows mNARX to be viewed as a series of physics-informed autoregressive models that break down the full problem into more easily solvable subproblems.

This process of reducing nonlinearity in the problem by constructing a more informative exogenous inputs manifold, can both improve the accuracy of the final surrogate model, and reduce the need for a large training dataset, without compromising prediction accuracy. 
Therefore, mNARX is ideal for situations where data is scarce, but prior information on the physics of the system is available.
In particular, mNARX can predict stably over long time periods, because all the components of the input manifold depend on the original exogenous input either explicitly or implicitly.

\subsubsection{Reduction of non-temporal coordinates}\label{sec:dimensionality_reduction}
NARX models require input data over multiple time steps, which often leads to high-dimensional regression problems, even when the raw exogenous input of the system is of moderate dimensionality. 
This can result in the \emph{curse of dimensionality}, which refers to the growth of model complexity with an increasing number of input features. 
The complexity of the model not only slows down its training process, but it also increases the risk of overfitting and obtaining a model that generalizes worse than a simpler one. 
To mitigate this issue, dimensionality reduction techniques can be applied to the original input to reduce its dimensionality, prior to the construction of the the exogenous input manifold (Section~\ref{sec:auxiliary_quantities}) and design matrix (Section~\ref{sec:ar_modelling_fitting}).

\begin{figure}
	\centering
	\includegraphics[width=8cm]{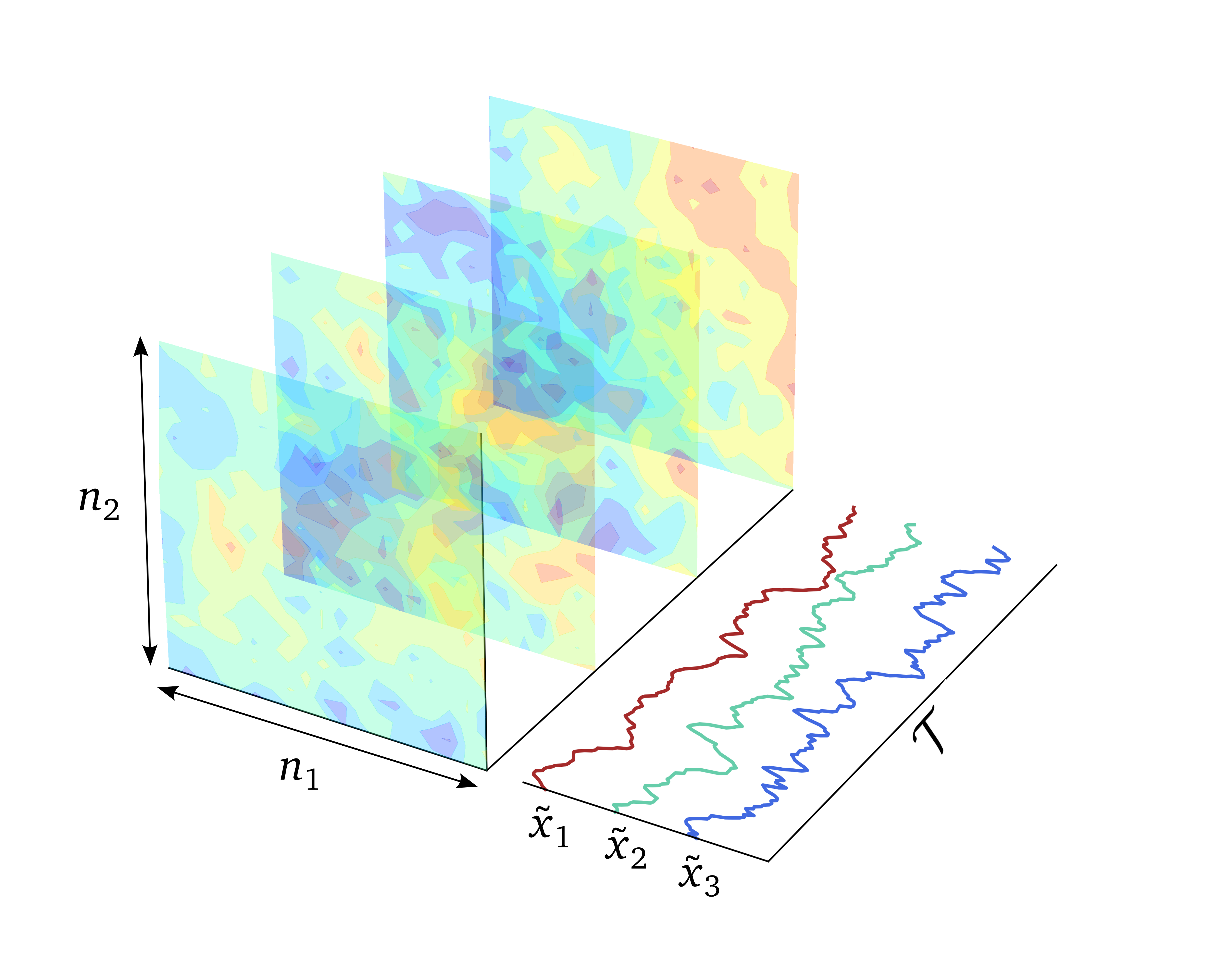}
	\caption{Illustration showing the creation of a small set of time-dependent features $\tilde{x}_i(t) \in \mathbb{R}$, $i=1,\dots,m$, from the original high-dimensional input $\vec{x}(t) \in \mathbb{R}^{M}$ where $M = n_1 \times n_2$ and $m=3$.}
	\label{fig:dim_reduction}
\end{figure}

In this paper, we focus on compressing the raw exogenous input $\vec{x}$ along its non-temporal coordinates using a transform $\mathcal{G}$, to obtain a lower dimensional representation $\tilde{\vec{x}}$:
\begin{equation}\label{eq:dim_reduction}
	\tilde{\vec{x}} = \mathcal{G}(\vec{x}),
\end{equation}
where $\tilde{\vec{x}} \in \mathbb{R}^{N \times m}$, $\vec{x} \in \mathbb{R}^{N \times M}$ and ideally $m \ll M$.
We do not compress the input in the time domain as the aim is to still model the system in its original time scale.
This approach allows us to keep Eq.~\eqref{eq:NARX}, and to use $\tilde{\vec{x}}$ instead of the raw $\vec{x}$ as part of the exogenous input manifold.
This concept is illustrated in Fig.~\ref{fig:dim_reduction}, which shows the reduction of an input with high spatial dimensionality to a small set of time-dependent features.
Note that this compression step is compatible with the construction of auxiliary quantities as shown in Section~\ref{sec:auxiliary_quantities}, Eq.~\eqref{eq:aux_quantity}, where we now replace $\vec{x}(\mathcal{T} \le t)$ with $\tilde{ \vec{x}}(\mathcal{T} \le t)$.

The mapping from the raw input data to the compressed input data can be performed using various techniques, such as auto-encoders \cite{rumelhart_1986}, principal component analysis (PCA) \cite{pearson_1901} and discrete cosine transform (DCT) \cite{ahmed_1974}, but also non-invertible and nonlinear techniques such as Isomap \cite{tenenbau_2000} or Kernel PCA \cite{goos_1997}.

Nonlinear transforms can often capture nonlinear relationships within the data better and preserve more of the local structure.
However, they tend to be computationally more complex than linear methods, and can be sensitive to the choice of their hyperparameters \cite{lee_2007}.

In some cases, the choice of a suitable dimensionality reduction technique can be straightforward. 
Typical examples include system responses that only depend on average properties of the raw inputs, such as strains or stresses on single nodes of components subject to spatially varying loads, which would be sensitive only to the first few spatial frequency modes. 
Other clear-cut cases are given by high dimensional inputs with extremely correlated components, such as is often the case with time-dependent boundary conditions on very fine finite-element meshes, which can be accurately parameterised with a small number of Karhunen-Lo\`eve decomposition eigenmodes \cite{shiryaev_2016}, or principal component analysis components.

In the end, the optimal choice of a dimensionality reduction technique depends on the specific problem being addressed, as well as on the characteristics of the input data. 
In general, it may be necessary to evaluate the performance of various techniques to determine the best approach for a given mNARX problem, e.g. as demonstrated in \cite{lataniotis_2019}.

\subsection{The mNARX algorithm}\label{sec:algorithm}\label{sec:polynomial_narx}
In summary, the mNARX algorithm comprises three main steps, for both its training phase on a given experimental design, and for its prediction stage as a surrogate model on unseen exogenous inputs.
The first step in the training phase involves optional data preprocessing, which includes operations such as scaling or interpolating the data in the time domain, as listed in Algorithm~\ref{alg:mnarx_training}.
Moving on to the second step, as discussed in Section~\ref{sec:auxiliary_quantities}, we distinguish between two types of auxiliary quantities: intermediate outputs and directly transformed features. The outputs of the direct transformation are not part of the experimental design data but can be computed from it. For example, this can involve using a dimensionality reduction method to obtain a lower-dimensional set of inputs for further computations. In contrast, the intermediate outputs are already present in the experimental design, and we need to calculate or train the corresponding functions $\mathcal{F}_i$ so that they will be available when making predictions on new, unseen data. 
It is worth noting that these intermediate outputs are initialized with the true data from the experimental design ($\vec{\beta}_i^\text{ED}$).
The third and final step involves training the auto-regressive surrogate model using the selected manifold features $\vec{z}_1, \dots, \vec{z}_n$.

\begin{algorithm}
	\caption{mNARX training algorithm}\label{alg:mnarx_training}
	\begin{algorithmic}
		\State \textbf{1. Input preprocessing} (time interpolation, scaling, etc.)
		\\
		\State \textbf{2. Training of the auxiliary functions ${\mathcal{F}_i}$ and construction of experimental design manifold}
		\For{$i \in \{1, 2, \dots, n\}$}    
		\If{ $\mathcal{F}_i(\bullet)$ is an intermediate output}
		\State train $\mathcal{F}_i$ on $\{{\vec{x}}^\text{ED}, \vec{z}_{1}^\text{ED}, \dots, \vec{z}_{i}^\text{ED}\}$
		\Else
		\State $z_i^\text{ED}(t) = \mathcal{F}_i({\vec{x}}^\text{ED}(\mathcal{T} \le t), \vec{z}_{<i}^\text{ED}(\mathcal{T} \le t), \vec{z}_{i}^\text{ED}(\mathcal{T} <t), \vec{\beta}^\text{ED}_i)$
		\EndIf
		\EndFor
		\\
		\State \textbf{3. Training of the final surrogate}
		\State train $\hat{\mathcal{M}}$ on $\{{\vec{x}}^\text{ED}, \vec{z}^\text{ED}_{1}, \dots, \vec{z}^\text{ED}_{n}, y\}$
	\end{algorithmic}
\end{algorithm}

When predicting new, unseen data, as shown in Algorithm~\ref{alg:mnarx_prediction}, we also begin with a preprocessing step, similar to the training algorithm.
In the second step, the prediction process differs from the training step in that all functions $\mathcal{F}_i$ are sequentially evaluated to generate predictions of the auxiliary variables $\hat{\vec{z}}_i$. However, during prediction, the initialization values $\vec{\beta}_i$ are typically unavailable and therefore must be set to sensible values, as mentioned in Section~\ref{sec:ar_modelling_prediction}.
In the last and final step, the final surrogate model is evaluated using all the auxiliary quantities that form the exogenous input manifold, starting from initial conditions $\vec{\beta}$.

\begin{algorithm}
	\caption{mNARX prediction algorithm}\label{alg:mnarx_prediction}
	\begin{algorithmic}
		\State \textbf{1. Input preprocessing} (time interpolation, scaling, etc.)
		\\
		\State \textbf{2. Calculation of the manifold components $\hat z_i(t)$ with the calibrated ${\mathcal{F}_i}$}
		\For{$i \in \{1, 2, \dots, n\}$}    
		\State $\hat{z}_i(t) = \mathcal{F}_i({\vec{x}}(\mathcal{T} \le t), \hat{\vec{z}}_{<i}(\mathcal{T} \le t), \hat{\vec{z}}_{i}(\mathcal{T} <t), \vec{\beta}_i)$
		\EndFor
		\\
		\State \textbf{3. Evaluation of the final surrogate}
		\State $\hat y(t) = \hat{\mathcal{M}}^\text{}({\vec{x}}(\mathcal{T} \le t), \hat{\vec{z}}_{1}(\mathcal{T} \le t), \dots, \hat{\vec{z}}_{n}(\mathcal{T} \le t), \hat{\vec{y}}(\mathcal{T} <t), \vec{\beta})$
	\end{algorithmic}
\end{algorithm}

In the subsequent applications in Section~\ref{sec:applications}, we use Polynomial NARX models as the auto-regressive surrogates. Due to their simple parametrization, efficient construction and fast evaluation, they are well-suited for use within the mNARX algorithm, which typically requires fitting multiple NARX models during the construction of the input manifold. They have also proven to be powerful surrogates in many applications \cite{mai_2016, spiridonakos_2015}. However, other popular and suitable alternatives include neural network-based models \cite{li_2021, song_2022} or models based on Gaussian process modelling \cite{koziel_2014, kocijan_2012} or support vector regression \cite{rankovic_2014}.

Polynomial NARX models are a special case of the ARX models introduced in Section~\ref{sec:ar_modelling}. 
For a given time instant $t$, they approximate $y(t)$ as a sum of monomials formed from the input vector $\vec{\varphi}(t)$ (see Eq.~\eqref{eq:exo_input}). 
We denote such a monomial $\mathcal{P}$ as
\begin{equation}\label{eq:polynomial}
	\mathcal{P}_{\vec{\alpha}}(\vec{\varphi}(t)) = \prod_{i=1}^{M_{\varphi}} \vec{\varphi}_i(t)^{\alpha_i},
\end{equation}
with $\vec{\varphi}_i(t)$ denoting the $i^\text{th}$ element of the vector $\vec{\varphi}(t)$ and $\vec{\alpha} \in \mathbb{N}^{M_\varphi}$ being an integer multi-index that defines the degree of the monomial of total degree $||\vec{\alpha}||_1 = \sum_{i=1}^{M_\varphi} \alpha_i$. 
We truncate the multi-index to a finite set 
$\vec{\alpha}  \in \mathcal{A}^{M_\varphi, d, r} : \left\{ (||\vec{\alpha}||_1 \le d) \cap (||\vec{\alpha}||_0 \le r)\right\}$, 
where $d$ is the maximum allowed polynomial degree, and  $r$ constrains the maximum interaction order  
$||\alpha||_0 = \sum_{i=1}^{M_{\varphi}} \mathbbm{1}_{\{\alpha_i > 0\}}$.
Using $\mathcal{A}$ as shorthand notation for the multi-index domain $\mathcal{A}^{M_\varphi, d, r}$ and given Eq.~\eqref{eq:polynomial} we can rewrite the general ARX formulation from Eq.~\eqref{eq:NARX} as 
\begin{equation}\label{eq:polynimial_NARX}
	y(t) = \sum_{\vec{\alpha}  \in \mathcal{A}} c_{\vec{\alpha}}  \mathcal{P}_{\vec{\alpha}}(\vec{\varphi}(t)),
\end{equation}
where we now express $ y(t)$ as a sum of all monomials weighted by a set of real-valued coefficients $c_{\vec{\alpha}}$.
For brevity, from now on we will denote the multivariate polynomial basis $\mathcal{P}_{\vec{\alpha}}(\vec{\varphi}(t))$ as $\mathcal{P}_{t}$. Further, we will use the super script $\mathcal{P}_{t}^{(i)}$ to indicate that $\mathcal{P}_{t}$ belongs to the $i^\text{th}$ realization in the experimental design with size $N_\text{ED}$. 
This allows us to define a regression matrix $\vec{\Psi}$ comprising the $\mathcal{P}_{t}$ of all time steps contained in the experimental design:
\begin{equation}
	\vec{\Psi} = 
	\{{\mathcal{P}_0^{(1)}}^\top, 
	\dots, 
	{\mathcal{P}_{t_\text{max}}^{(1)}}^\top, 
	\dots, 
	{\mathcal{P}_0^{(N_\text{ED})}}^\top, 
	\dots, 
	{\mathcal{P}_{t_\text{max}}^{(N_\text{ED})}}^\top
	\}^\top.
\end{equation} 
Analogously, we define an output vector $\vec{y}$ corresponding to the input matrix $\vec{\Psi}$ as
\begin{equation}
	\vec{y} = 
	\{y_0^{(1)}, 
	\dots, 
	y_{t_\text{max}}^{(1)}, 
	\dots, 
	y_0^{(N_\text{ED})}, 
	\dots, 
	y_{t_\text{max}}^{(N_\text{ED})}
	\}^\top.
\end{equation} 
Note that, despite the initial temporal coherence of the exogenous input and output, the rows of the design matrix $\vec{\Phi}$ and hence also of $\vec{\Psi}$ do not need to follow any temporal order, as explained in Section~\ref{sec:ar_modelling}.
Therefore, determining the set of model coefficients reduces to a linear regression problem $\vec{y} = \vec{\Psi}\vec{c} + \varepsilon$ in which $\vec{c}=\{c_\alpha, \vec{\alpha} \in \mathcal{A}\}$ collects the polynomial coefficients and $\varepsilon$ is the residual error. 
We compute these coefficients by means of ordinary least squares minimization, which is computationally efficient:
\begin{equation}
	\hat{\vec{c}} = \mathop{\arg\min}_{\vec{c} \in \mathbb{R}^{|\mathcal{A}|}} || \vec{y} - \vec{\Psi} \vec{c} ||^2
\end{equation}

\section{Applications}\label{sec:applications}

\subsection{Coupled spring-mass system}\label{sec:spring_mass_system}
\subsubsection{Problem statement}\label{sec:spring_mass_system_problem_statement}
\begin{figure}[htb]
	\centering
	\includegraphics[width=0.22\textwidth]{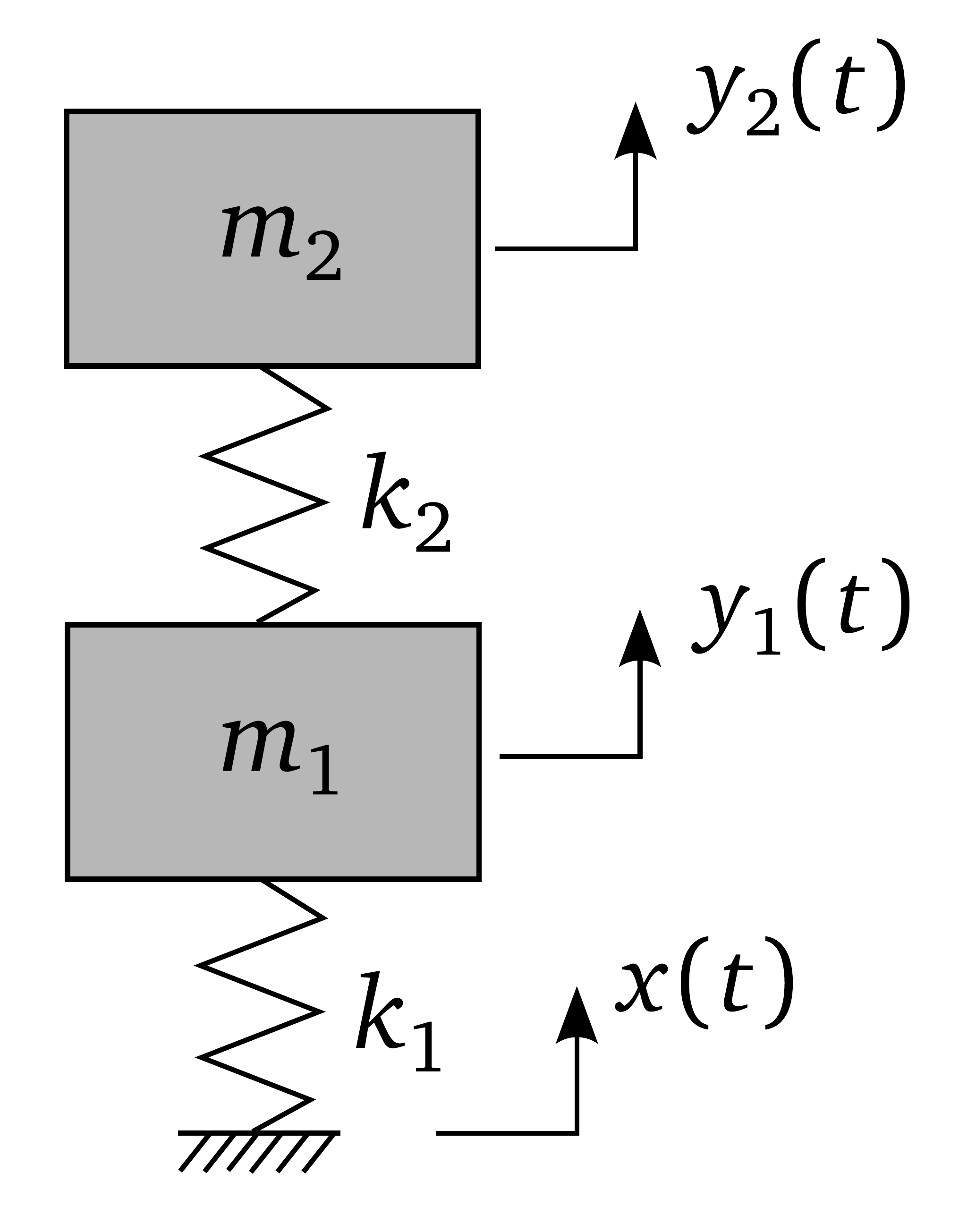}
	\caption{Coupled spring-mass system}
	\label{fig:mass_spring_system}
\end{figure}

In our first application, we consider the coupled spring-mass system sketched in Fig.~\ref{fig:mass_spring_system}, which consists of two masses $m_1$ and $m_2$ connected by a linear spring with stiffness $k_2$.
The lower mass is fixed to the ground by a spring with stiffness $k_1$. 

The system is described by a system of ordinary differential equations with two degrees of freedom:
\begin{equation}\label{eq:quarter_car}
	\begin{cases}
		& m_2 \, \ddot{y}_2(t) = -k_2\left(y_2(t)-y_1(t)\right), \\
		& m_1 \, \ddot{y}_1(t) = k_2\left(y_2(t)-y_1(t)\right) + k_1\bigl(x(t)-y_1(t) \bigl).
	\end{cases}
\end{equation}
The numerical values for the system parameters are listed in Tab.~\ref{tab:mass_spring_system}. 
The upper mass $m_2$ is about two orders of magnitude smaller than the lower mass $m_1$, while the two ratios $\frac{k_1}{m_1}$ and $\frac{k_2}{m_2}$ are similar. 
Consequently, the displacement of the upper mass $y_2$ is strongly dependent on the displacement of the lower mass $y_1$, while the displacement of the lower mass is largely unaffected by the upper mass.

\begin{table}[htb]
	\centering
	\caption{Coupled spring-mass system -- System parameters}\label{tab:mass_spring_system}
	\begin{tabular}{@{}lcc@{}}
		\toprule
		Parameter & Unit & Value \\ \midrule
		Spring stiffness $k_1$ & N/mm & 10{,}000 \\
		Spring stiffness $k_2$ & N/mm & 100 \\
		Mass $m_1$ & kg & 300 \\ 
		Mass $m_2$ & kg & 2 \\ \bottomrule
	\end{tabular}
\end{table}

We study these two displacements under a random excitation 
\begin{equation}\label{eq:mass_spring_system_exo_input}
	x(t) = \frac{1}{N_\omega}\sum_{i=1}^{N_\omega} A_i \sin \left( 2\pi B_i t + C_i\right)
\end{equation}
acting on $m_1$ via the lower spring.
The excitation is the arithmetic mean of $N_\omega$ sinusoidal terms.
The number of terms is modelled as a discrete uniform random variable $N_\omega$, with $\Prob{N_\omega=i}_{i=1,\dots, 10} = \frac{1}{10}$. 
Each of the terms has random amplitude $A_i$ and frequency $B_i$, both of which follow a uniform distribution $\sim \mathcal{U}(-1, 1)$.
The phase $C_i$ also follows a uniform distribution, $C_i \sim \mathcal{U}(-\pi, \pi)$.
Besides being easily interpretable, Eq.~\eqref{eq:mass_spring_system_exo_input} allows us to generate monochromatic as well as frequency-rich excitations.

On this spring-mass system we compare mNARX and classical NARX models, in terms of their ability to predict both displacements $y_1(t)$ and $y_2(t)$ from an unseen excitation $x(t)$. 
Both approaches are trained and validated on 30-second realizations of the system starting at rest and sampled with a time step $\delta t=0.01$~s. 
Because the number of samples ($N=3001$) in each realization and the dimensionality of the exogenous input ($M=1$) is low, training of the models is performed on all samples from the design matrix (see Section~\ref{sec:ar_modelling_fitting}) built from $N_\text{ED} = 5$ random simulations of the system.
Validation is performed on a large set of $N_\text{val} = 10{,}000$ out-of-sample simulations.

\subsubsection{mNARX and NARX configuration}\label{sec:model_structure}
In the NARX approach, we build two polynomial NARX models to predict $y_1(t)$ and $y_2(t)$ independently as a function of the exogenous excitation $x(t)$. The model structures including the truncation scheme, the included lags and the total number of polynomial coefficients are listed in Tab.~\ref{tab:model_configuration}. 

With mNARX, we first predict $y_1(t)$ as a function of $x(t)$, identically to the NARX approach.
We then predict $y_2(t)$ with two exogenous inputs, namely $x(t)$ and the prediction $\hat y_1(t)$. This allows us to keep the total number of model coefficients smaller than in the NARX approach, as shown in Tab.~\ref{tab:model_configuration}, and helps avoid overfitting on the small training set.

The model configuration has been selected in an iterative process of experimentation and refinement for each of the surrogates. 
In this process, we tried to minimize the model complexity while achieving good predictive accuracy.

For both approaches we initialize the NARX models with the first few time steps of the true output as detailed in Section~\ref{sec:ar_modelling_prediction}. 
Although not necessary, this allows us to better validate and compare the two methods because the temporal evolution of the system is sensitive to its initialization.

\begin{table}[htb]
	\centering
	\caption{Coupled spring-mass system -- Model configurations of the mNARX and classic NARX approach}
	\begin{tabular}{@{}lccc@{}}
		\toprule
		& \multicolumn{1}{c}{$y_1$}  & \multicolumn{2}{c}{$y_2$}   \\ \midrule
		& \multicolumn{1}{c}{\textbf{NARX/mNARX}} & \multicolumn{1}{c}{\textbf{NARX}} & \multicolumn{1}{c}{\textbf{mNARX}} \\
		Exogenous inputs           & $x(t)$                    & $x(t)$                      & $x(t), \yhat_1(t)$  \\
		Maximum polynomial degree  & 1                      & 1                        & 1  \\
		Interaction order          & 1                      & 1                        & 1  \\
		Auto-regressive lags      & $\{\dt, 2\dt, 3\dt\}$   & $\{\dt, \dots, 20\dt\}$   & $\{\dt, 2\dt, 3\dt\}$ \\
		Exogenous input lags      & $\{0, \dt, 2\dt\}$ &   $\{0, \dots, 3\dt\}$   & $\{0, \dt\}$, $\{0, \dt, 2\dt\}$ \\
		Number of coefficients     & 6                      & 24                        & 8 \\ \bottomrule
	\end{tabular}
	\label{tab:model_configuration}
\end{table}

\subsubsection{Results}\label{sec:quarter_car_results}
\begin{figure}[htb]
	\centering
	\begin{subfigure}[b]{0.49\textwidth}
		\centering
		\includegraphics[width=\textwidth]{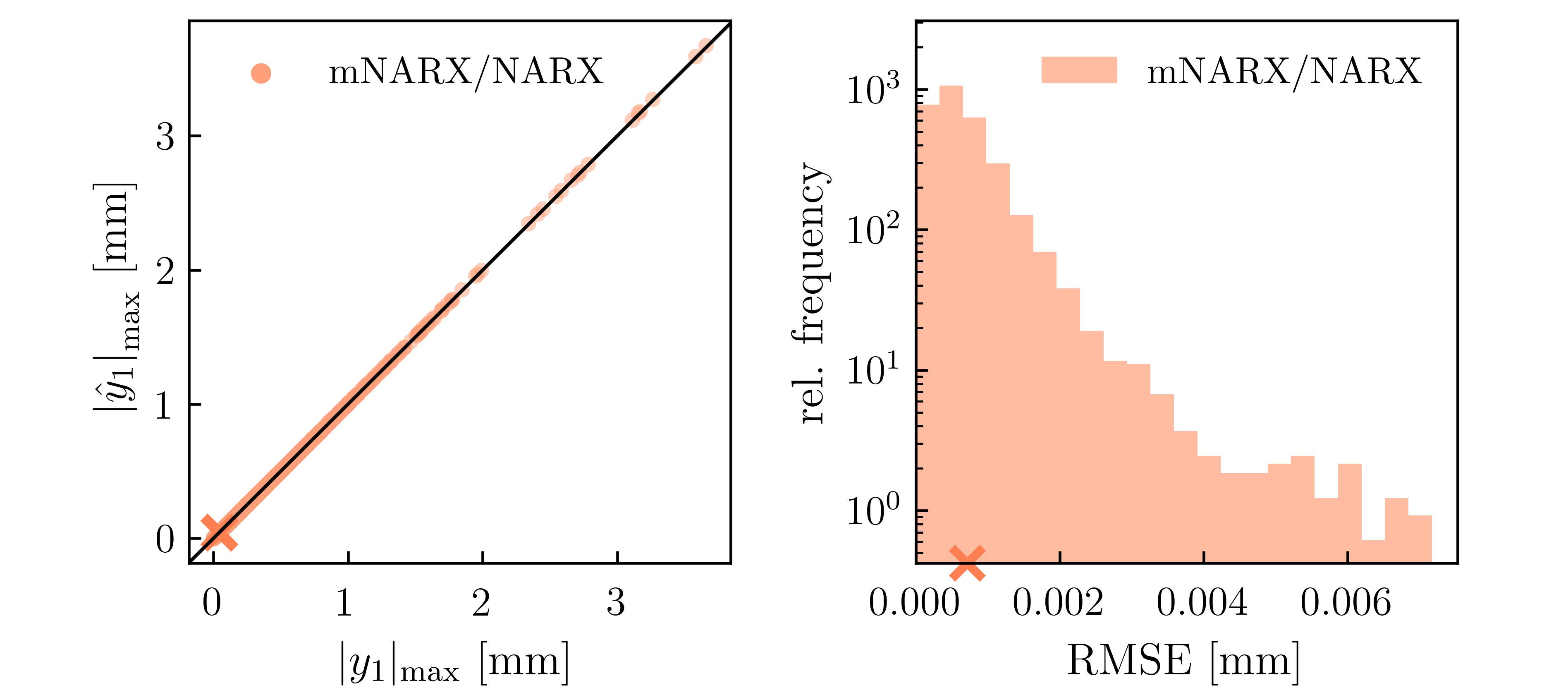}
	\end{subfigure}
	\begin{subfigure}[b]{0.49\textwidth}
		\centering
		\includegraphics[width=\textwidth]{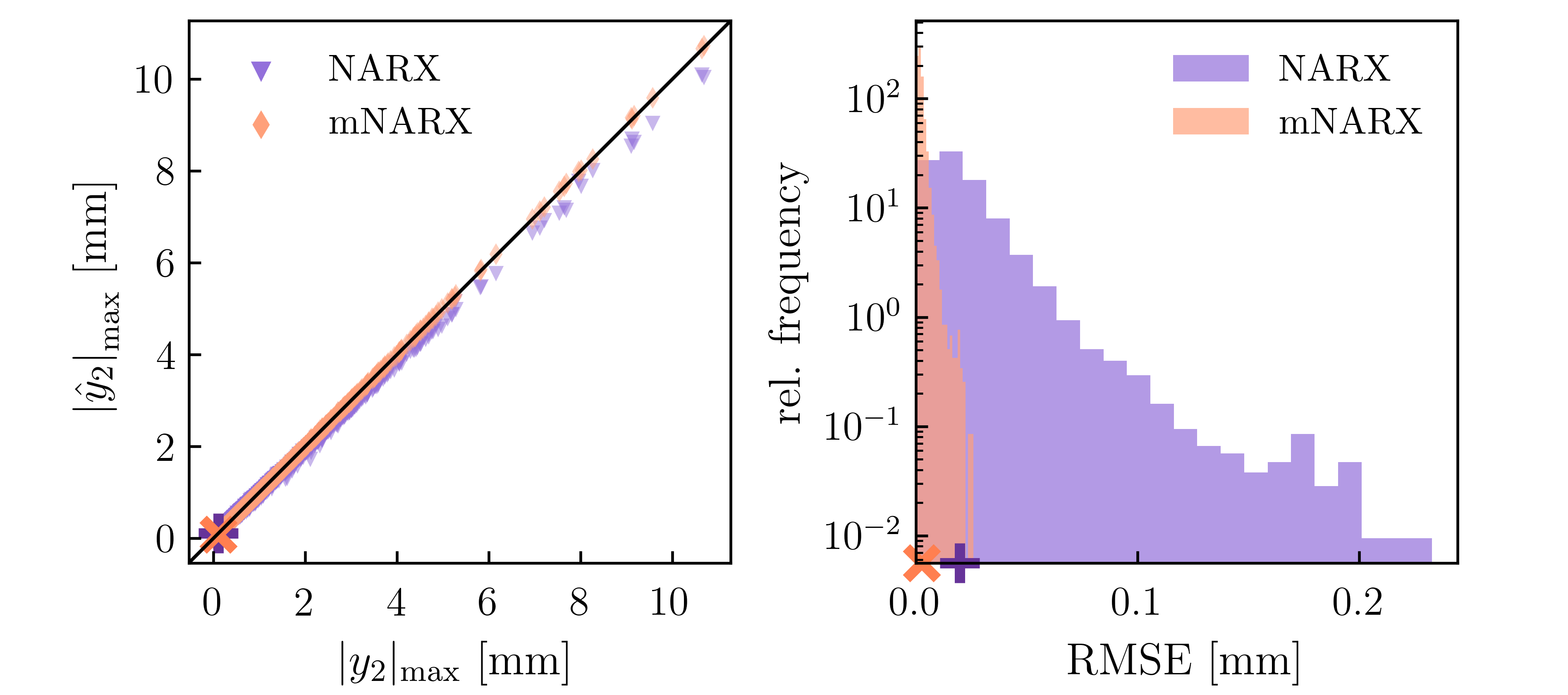}
	\end{subfigure}
	
	\begin{subfigure}[b]{0.49\textwidth}
		\centering
		\includegraphics[width=\textwidth]{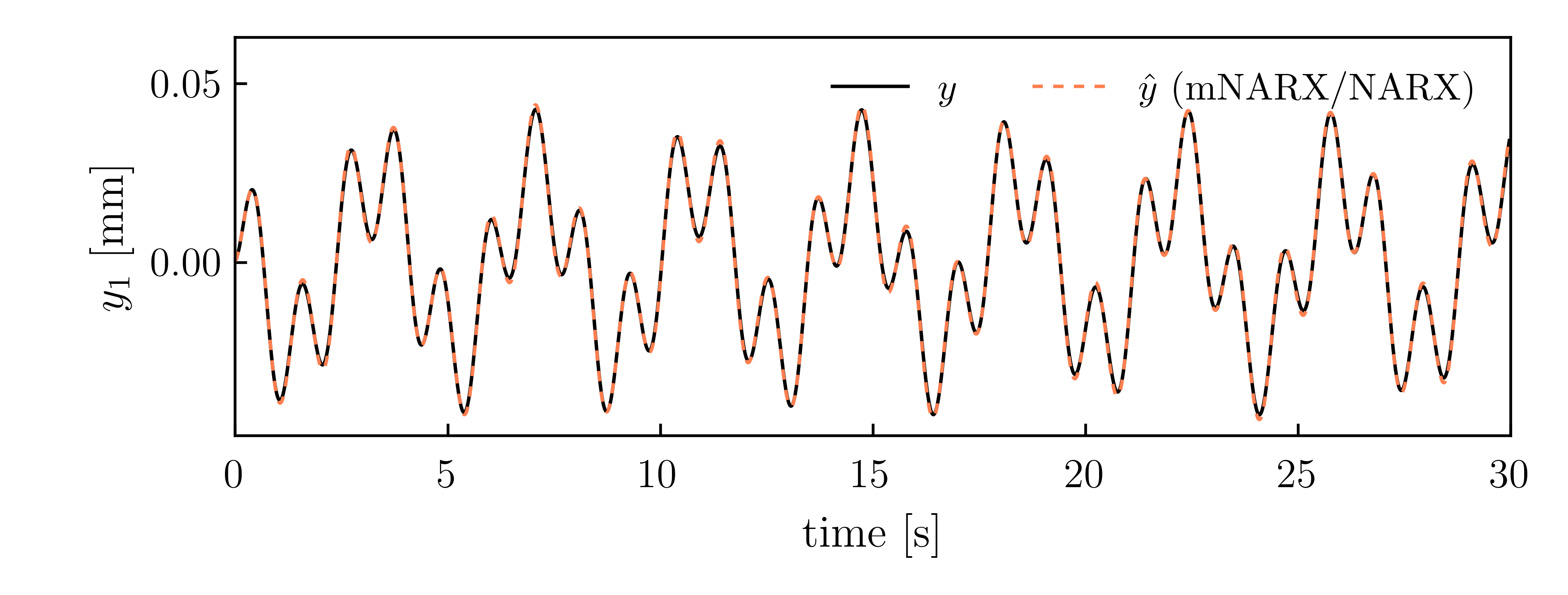}
		\caption{Lower mass displacement $y_1$}
	\end{subfigure}
	\begin{subfigure}[b]{0.49\textwidth}
		\centering
		\includegraphics[width=\textwidth]{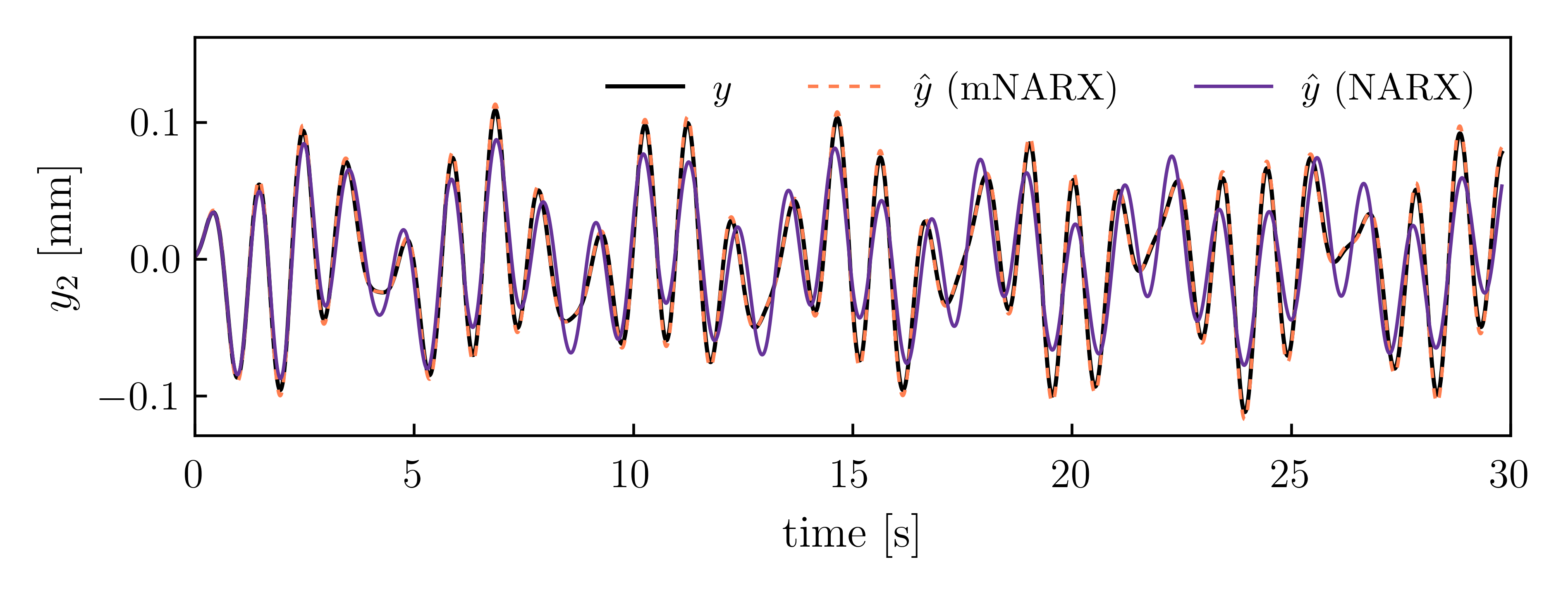}
		\caption{Upper mass displacement $y_2$}
	\end{subfigure}
	
	\caption{Coupled spring-mass system results.            
		(a)~
		(Top left) Predicted maximum displacement of the lower mass $|\hat y_1|_{max}$ vs. the true displacement $|y_1|_{max}$ identical by construction for the mNARX and NARX surrogates.
		(Top right) Histogram of the root-mean-squared error (RMSE) for the prediction of the displacement $y_1$. 
		(Bottom) Exemplary trace of the lower mass displacement. The true displacement is depicted as a solid black line while the prediction is shown as a dashed blue line. The trace corresponds to the point and RMSE marked by an orange cross in the scatter plot and histogram.
		(b)~
		(Top left) Predicted vs. true maximum displacement of the upper mass $y_2$. The results for the NARX approach are shown in purple and the mNARX results are shown in orange.
		(Top right) Histogram of the RMSE for the prediction of the displacement $y_2$. 
		(Bottom) Exemplary trace of the upper mass displacement. The true displacement is depicted in black, the mNARX prediction as a dashed orange line and the NARX prediction in purple. The traces correspond to the orange and purple crosses in the scatter plot and histogram.
	}
	\label{fig:mass_spring_system_results}
\end{figure}

\newcommand{\qcmfig}[1]{Fig.~\ref{fig:mass_spring_system_results}{#1}}
The performance of the NARX and mNARX method on the mass-spring system validation set are shown in \qcmfig{}. The results for the prediction of the lower mass displacement are shown in \qcmfig{a} and the result for the upper mass displacement in \qcmfig{b}.

In the top left panel of \qcmfig{a} we provide a visual comparison between the true absolute peak displacements of the lower mass $|y_1|_{max}$ and the predicted one $|\yhat_1|_{max}$. 
The root-mean-squared error (RMSE) on this quantity is shown in the top right panel.
The full trace of the realization marked by an orange cross is displayed in bottom panel. 
These plots show that a linear model with only 6 terms predicts well the peak displacement even for the extremely low and high values.
Further, the model prediction is accurate and stable over the full 30 seconds duration of the simulation.
The good results from the scatter plot are matched by their very low RMSEs shown in the histogram. 

The comparison of mNARX and the standard NARX approach on $y_2$ is given in \qcmfig{b}.
We depict the mNARX results in orange and the NARX results in purple.
In the top left panel we see generally good agreement of the true and predicted peak displacement for both methods. 
However, for the NARX approach there is a trend toward underestimating the true displacement especially when the peak displacement becomes large.
The inferior performance of the NARX approach compared to mNARX becomes even more apparent when looking at the RMSE shown in the top right panel. 
The accuracy of the NARX approach is considerably lower and less consistent for rare events as it can be seen from the long tail of the histogram (note that the ordinate is in log scale).
In the bottom panel it can be seen that the standard NARX surrogate is not capable of modelling the system response accurately. 
It shows a clear mismatch in magnitude and phase. 
Even though the mNARX model consists of only 8 terms compared to the 24 terms of the standard NARX model, its prediction closely follows the true system response. 
Furthermore, the prediction $\hat y_2$ is stable over the entire 30 seconds and virtually no error accumulates over time.

\subsection{Wind turbine simulation}\label{sec:windturbine_simulation}

\subsubsection{Problem statement}\label{sec:windturbine_problem_statement}
\begin{figure}[htb]
	\centering
	\includegraphics[width=0.32\textwidth]{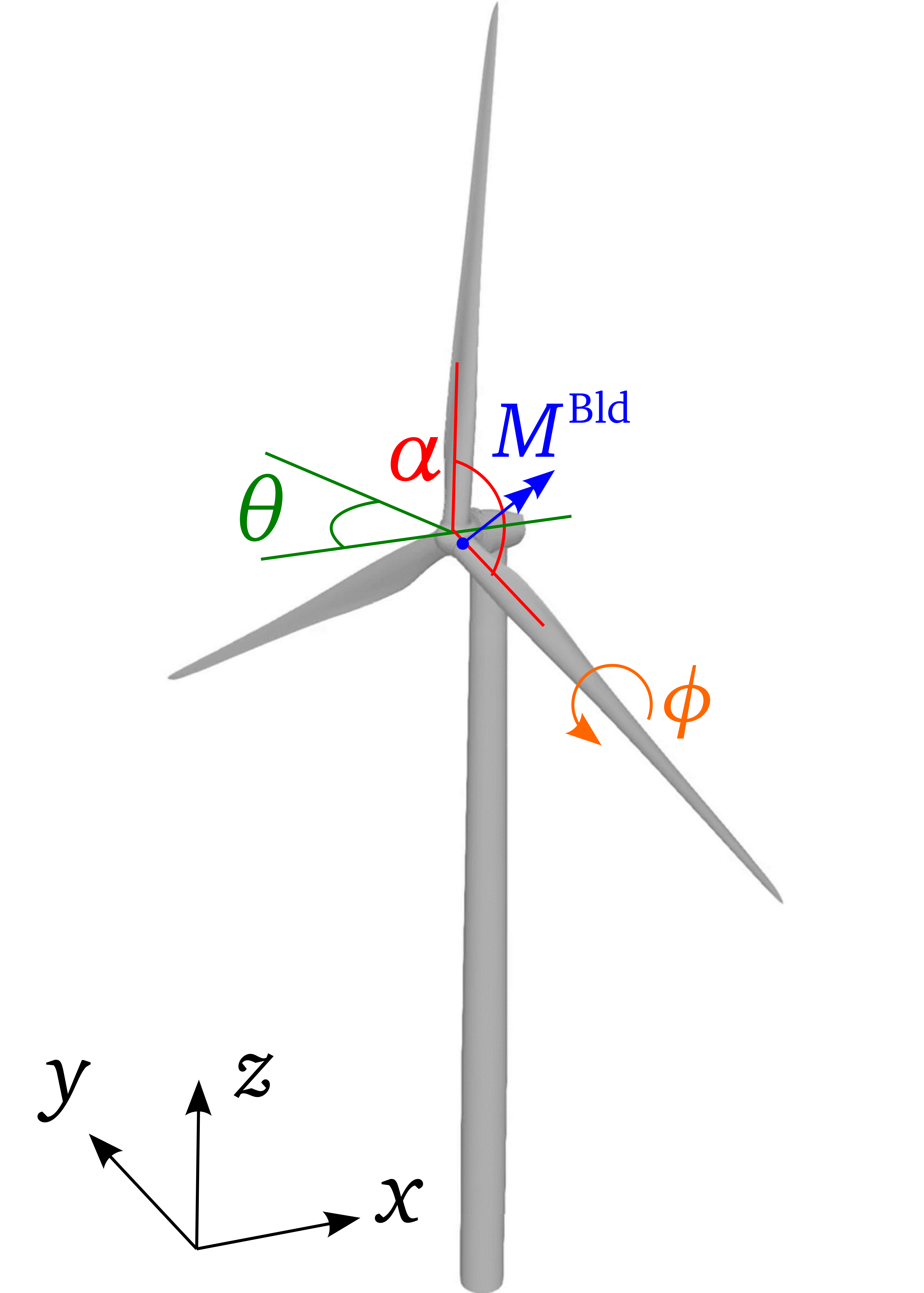}
	\caption{
		Sketch of an onshore wind turbine. 
		The following degrees of freedom and loads are annotated: 
		rotor azimuth angle $\alpha$ (red),
		blade pitch angle $\phi$ (orange),
		yaw angle $\theta$ (green),
		flapwise blade root bending moment $M^\text{Bld}$ (blue).
	}
	\label{fig:wind_turbine}
\end{figure}

After demonstrating mNARX on an analytical case study with low exogenous input dimensionality, we now use mNARX to emulate a realistic engineering scenario: an aero-servo-elastic (ASE) wind turbine simulator.
The input to the simulator is a four-dimensional, temporally-coherent random field that represents the wind speeds across the 2D-area spanned by the turbine rotor:
\begin{equation}
	\vec{v}: \mathcal{T} \to \mathbb{R}^{\nu_w \times \nu_y \times \nu_z},
\end{equation}
called a \textit{turbulence box}. 
At every time instant on a discrete time axis $\mathcal{T}$, the turbulence box is described by $\nu_w=3$ wind speed components (longitudinal $x$, transversal $y$ and vertical direction $z$) at every one of the $\nu_y \times \nu_z$ spatial grid points. 
Assuming a typical spatial discretization of $\mathcal{O}(10^1)$ in either direction, this results in a total spatial dimensionality of $\mathcal{O}(10^{2-3})$.
The outputs of the ASE simulator are multiple univariate time series 
\begin{equation}
	f_i: \mathcal{T} \to \mathbb{R},
\end{equation}
such as the evolution of the blade- or tower- internal forces (e.g. bending moments), blade pitch, or the instant power production. 
Our goal is to construct an mNARX surrogate $\hat{\mathcal{M}}$ that predicts the quantity $f_i$ at every time step $t$ solely based on the wind speeds up to and including time $t$:
\begin{equation}
	f_i(t) = \hat{\mathcal{M}}({\vec{v}(\mathcal{T} \le t)}).
\end{equation}
Besides the high-dimensional exogenous wind input, additional complexity arises in the ASE simulations from the highly nonlinear and non-differentiable relationship between input wind and some of the output quantities. 
This nonlinearity and non-differentiability are introduced by the turbine controller system which manipulates multiple degrees of freedom (DOFs) of the turbine blades and the nacelle. 
Moreover, the problem is further complicated by the fact that many output quantities depend on the orientation of the rotor blades due to gravity and wind shear.
An illustration of an onshore wind turbine with its DOFs is given in Fig.~\ref{fig:wind_turbine}. 

\subsubsection{Computational model}\label{sec:simulation_setup}
In this study, we conduct ASE simulations on the well known reference 5MW NREL baseline onshore wind turbine \cite{nrel_onshore_2009}, using the NREL ROSCO (Reference Open-Source Controller) \cite{rosco_2022} and the open-source low-fidelity ASE simulator OpenFAST \cite{openfast_2021}, with a time step of 0.00625~s. 
A summary of the turbine specifications can be found in Tab.~\ref{tab:turbine_specs}.

\begin{table}[htb]
	\centering
	\caption{NREL 5MW reference turbine specifications}
	\begin{tabular}{@{}ll@{}}
		\toprule
		Category                          & Specification                   \\ \midrule
		Rated power                       & 5 MW                               \\
		Rotor orientation, configuration  & upwind, 3 blades                   \\
		Control                           & variable speed, collective pitch   \\
		Drivetrain                        & high speed, multiple-stage gearbox \\
		Rotor, hub diameter               & 126 m, 3 m                         \\
		Hub Height                        & 90 m                               \\
		Cut-in, rated, cut-out wind speed & 3 m/s, 11.4 m/s, 25 m/s            \\
		Cut-in, rated rotor speed         & 6.9 rpm, 12.1 rpm                   \\ \bottomrule                 
	\end{tabular}
	\label{tab:turbine_specs}
\end{table}

To generate the input turbulence boxes, we use the open-source stochastic turbulence box generator TurbSim \cite{turbsim_2009}. 
TurbSim generates turbulence boxes from random realizations of a set of scalar environmental condition summaries. 
These parameters are modelled as random variables according to the normal turbulence model for wind turbine class B I, as defined by the IEC 61400-1 design standard for onshore wind turbines \cite{iec_2005}. 

The standard defines two random parameters: 
\begin{itemize}
	\item the \textit{reference wind speed} $V_\text{hub}$, which follows a Rayleigh distribution and is defined as the mean longitudinal wind speed at hub height over the total duration of the generated turbulence box
	\item and the \textit{turbulence standard deviation} of the longitudinal wind speed at hub height, $\sigma_1$, which follows a lognormal distribution conditional on $V_\text{hub}$. 
\end{itemize}
The turbulence box is constructed using the Kaimal spectral model and an exponential coherence model, as suggested by the design standard. 
As a result, the turbulence box is coherent both in time and space for the longitudinal wind speed component but incoherent in space for the transverse and upwards wind speed components. 
Additionally, the three components of the wind speed differ in terms of their magnitude, with the longitudinal component being about one order of magnitude larger than the transverse and upwards ones. The longitudinal component $\vec{v_x}$ is also superimposed by a wind shear profile (varying with the altitude $z$) with a constant shear coefficient $\alpha=0.2$, according to the standard, resulting in generally higher wind speeds at higher altitudes. The air density is modelled as a constant with a value of $1{,}225$~kg/m$^3$.
For discretization, we use a spatial grid resolution of 19 by 19 ($\nu_y=\nu_z=19$) grid points, resulting in an exogenous input with a spatial dimensionality of 1{,}083. 
The wind speed is sampled at 20~Hz (time step of 0.05~s) while the ASE solver uses a time step of 6.25~ms. Thus the turbulence box is interpolated to this 8-time higher rate.  
The full set of parameters used for TurbSim is listed in Tab~\ref{tab:turbsim_specs}.

\begin{table}[htb]
	\centering
	\caption{TurbSim parameters (detailed distributions can be found in \cite{iec_2005})}
	\begin{tabular}{@{}lcc@{}}
		\toprule
		Parameter & Unit & Value \\ \midrule
		Reference wind speed  $V_\text{hub}$ & m/s & Rayleigh distribution \cite{iec_2005} \\
		Turbulence standard deviation $\sigma_1$ & m/s & Lognormal distribution conditional on $V_\text{hub}$ \cite{iec_2005} \\
		Wind shear $\alpha$ & - & $0.2$  \\
		Air density $\alpha$ & - & $1{,}225$~kg/m$^3$  \\
		Spatial discretization & - & $19 \cdot 19$ \\     
		Temporal discretization & s & $0.05$ \\ \bottomrule     
	\end{tabular}
	\label{tab:turbsim_specs}
\end{table}

\subsubsection{Output quantities of interest}\label{sec:quantities_of_interest}
In this study, we focus on two quantities of interest (QoI) for the design of wind turbines, namely the blade root bending moment (more specifically the \textit{flapwise moment} $M^\text{Bld}(t)$), and the generated power of the turbine $P(t)$. 
The accurate prediction of the flapwise bending moment is crucial for fatigue and ultimate limit state design, thus for the reliability of the turbine.
From expert knowledge, we know that these two QoIs depend both on the rotor speed $\omega(t)$ and the blade pitch $\phi(t)$ imposed by the controller as soon as the wind speed exceeds the rated wind speed (see Fig.\ref{fig:wind_turbine}). Thus $\omega(t)$ and $\phi(t)$ will be our auxiliary quantities.

As outlined in Tab.~\ref{tab:turbine_specs}, the 5MW NREL turbine has a rated wind speed of 11.4~m/s, at which it reaches its maximum power output of 5~MW and adjusts the blade pitch to maintain this level. 
To take advantage of this knowledge, we construct two separate mNARX surrogates, namely one for \textit{below-rated} wind speeds and another one for \textit{above-rated} wind speeds. 
We classify each turbulence box used for prediction and validation based on the reference wind speed $V_\text{hub}$. 
As this is a known characteristic of the turbulence box, the exogenous input, it is also available for the out-of-sample wind boxes in the validation set.

\subsubsection{Training and validation data}\label{sec:train_test_data}
In this study, we conducted a total of 1{,}050 ASE simulations, each with a duration of 12 minutes, using the computational model detailed in Section~\ref{sec:simulation_setup}. 
To ensure the validity of our data, we truncated the first two minutes of each simulation, which include the start-up phase of the turbine. 
These 1{,}050 simulations were divided into two regimes, with 600 simulations belonging to the below-rated wind speed regime and 450 to the above-rated wind speed regime. 
To ensure robustness, we randomly selected 100 simulations from each regime to serve as training datasets for each emulator, while the remaining simulations were used for validation.

\subsubsection{mNARX structure}\label{sec:mNARX_structure}
The high spatial dimensionality of the turbulence box ($M=3 \cdot 19 \cdot 19 = 1{,}083$) makes it infeasible to use it directly as an exogenous input to model any of the turbine response quantities. 
To reduce its dimensionality, we only keep the longitudinal wind speed component of the turbulence box, $\vec{v_x}$, as it is the most relevant in this context ($M=19 \cdot 19 = 361$) .
Additionally, the transverse and vertical wind speeds, which have no spatial coherence by construction (see Section~\ref{sec:simulation_setup}), can be considered as noise.
To further reduce the dimensionality of the data, each 2D slice $\vec{v}_x(t)$ of the turbulence box, characterized by pixels $v_x^{\kappa, \ell}(t)$, ${k, l = 1, \dots 19}$, is represented by its 2D discrete cosine transform (DCT) coefficients $\xi(t)$:
\begin{equation}\label{eq:discrete_cosine_transform}
	v_x^{\kappa, \ell}(t) = \sum_{i=0}^{n_i-1}\sum_{j=0}^{n_j-1} \xi_{i,j}(t)
	\cos\left[ \frac{\pi}{\nu_y} \left( \kappa+\frac{1}{2} \right)i \right]
	\cos\left[ \frac{\pi}{\nu_z} \left( \ell+\frac{1}{2} \right)j \right].
\end{equation}
Subsequently, we will refer to the coefficients $\xi(t)$ as the spatial modes of the input $v_x(t)$. They represent different spatial frequencies within each slice $\vec{v}_x$. Lower-frequency coefficients tend to capture the broader, smoother patterns in the image, while higher-frequency coefficients capture finer details.
The dimensionality of $\vec{v}_x$ is reduced by selecting $n_i$ and $n_j$ such that $n_i < \nu_y = 19$ and $n_j < \nu_z = 19$. 
For simplicity, we choose the same number of coefficients in the two spatial direction ($n_i=n_j$).
As the system response is mostly governed by the low spatial frequency components of the wind, we only keep the 2-5 coefficients corresponding to the lowest frequencies in each direction.
This reduces the spatial dimensionality of each trajectory $\vec{v}_x(t)$ by 1-2 orders of magnitude, from 361 to 4-25.

We use the time-dependent spectral coefficients $\vec{\xi}$ in Eq.~\eqref{eq:discrete_cosine_transform} as the exogenous input to build a first polynomial NARX model (see Section~\ref{sec:polynomial_narx}) to model the blade pitch $\phi(t)$ of the turbine:
\begin{equation}
	\hat\phi(t) = \hat{\mathcal{M}}_\phi^\text{}(\vec{\xi}_\phi(\mathcal{T} \le t)),
\end{equation}
where the subscript $\phi$ in $\vec{\xi}$ denotes that we only use a subset of $\vec{\xi}$ as the exogenous input to $\hat{\mathcal{M}}_\phi$. 
Modelling the blade pitch is a crucial first step as it determines the angle of attack and the amount of wind striking the blades, and therefore, how strongly the turbine responds to the wind. 
Because the ROSCO controller adjusts the blade pitch mostly based on the inflowing wind, using only $\vec{\xi}$ as an exogenous input is sufficient for this QoI.

When building the NARX model, the wind speed component $\vec{v}_x$ is generated at 20~Hz, but the simulator output has a sampling frequency of 160~Hz, so we upsample $\vec{\xi}$ to match the sampling frequency of the simulator. 
With 100 training simulations, each 600~s long, this results in a large amount of data, with a total of $\mathcal{O}(10^7)$ time steps, which would cause the regression matrix (see Section~\ref{sec:polynomial_narx}) in the linear regression step to become extremely large and prohibit using least-squares on ordinary computer hardware.
To avoid this issue, we use a random subset of the design matrix as described in Section~\ref{sec:ar_modelling_fitting} and the corresponding output vector to perform the least-square fitting of the polynomial NARX.
The number of random subsamples ranges between $10^5$ and $10^6$ depending on the number of model coefficients.
By doing this, we can ensure that the model is well-conditioned and not overfitting, while at the same time reducing the computational costs associated with training. 
Note that the resulting number of training samples can still be high relative to the number of regression coefficients. 
While it is possible to further reduce the number of samples without compromising the accuracy of the surrogates or choose a more complex model structure, our empirical iterative refinement process to determine the model configuration did not yield significant improvements in accuracy with increased numbers of lags or polynomial degrees.
Moreover, using fewer samples also does not provide a significant benefit, since the computational cost of fitting the model is already relatively low. 
We interpret this behaviour as a sign that, in this particular case study, prediction accuracy is limited either by the NARX structure, or by the choice of the manifold.

The final size of the random subset of samples used to train $\hat{\mathcal{M}}_\phi$ and the model configuration for both wind speed regimes can be found in Table~\ref{tab:blade_pitch_mNARX_config}.
Note that, due to changes in the turbine control system as discussed in Section~\ref{sec:quantities_of_interest}, the two model configurations differ.

\begin{table}[htb]
	\centering
	\caption{Wind turbine simulation -- Configuration of the NARX surrogate for the prediction of the blade pitch $\phi(t)$}
	\begin{tabular}{@{}lcc@{}}
		\toprule
		\multicolumn{1}{l}{\multirow{2}{*}{Exogenous inputs:}} & \multicolumn{2}{c}{$\vec{\xi}_{0,0}(t), \dots, \vec{\xi}_{2,2}(t)$}                      \\ \cmidrule(l){2-3} 
		\multicolumn{1}{l}{}                                   & Below-rated wind speed & Above-rated wind speed \\ \midrule
		Maximum polynomial degree                              & 3                     & 5                      \\
		Interaction order                                      & 1                     & 1                      \\
		Auto-regressive lags                                   & $\{1,160\}$           & $\{1,160\}$            \\
		Exogenous input lags                                   & $\{1\}$               & $\{1\}$                \\
		Number of coefficients                                 & 33                    & 55                      \\
		Number of training samples                             & $\mathcal{O}(10^6)$                   & $\mathcal{O}(10^6)$                     \\ \bottomrule
	\end{tabular}
	\label{tab:blade_pitch_mNARX_config}
\end{table}

In the second modelling step, we construct a NARX model to predict the rotor speed $\omega(t)$. 
The rotor speed is a crucial quantity in wind turbine operation, as it determines the power generation and the evolution of the rotor azimuth. 
The latter is strongly connected to the blade loads because of gravity and because a blade pointing upwards usually experiences higher wind speeds.
The rotor speed depends in turn on the blade pitch $\hat\phi(t)$ that we modelled in the first step. 
Therefore, we create an input manifold $\vec{\zeta}_\omega(t) = \{\vec{\xi}_\omega(t), \hat\phi(t)\}$ as the exogenous input for the NARX model. 
The NARX model is then represented by:
\begin{equation}
	\hat\omega(t) = \hat{\mathcal{M}}_\omega^\text{}(\vec{\zeta}_\omega(\mathcal{T} \le t)).
\end{equation}
The configuration of the NARX model and the number of random subsamples used for the least-square fit are listed in Tab.~\ref{tab:rotor_speed_mNARX_config}.
\begin{table}[htb]
	\centering
	\caption{Wind turbine simulation -- Configuration of the NARX surrogate for the prediction of the rotor speed $\omega(t)$}
	\begin{tabular}{@{}lcc@{}}
		\toprule
		\multicolumn{1}{l}{\multirow{2}{*}{Exogenous inputs:}} & \multicolumn{2}{c}{$\vec{\xi}_{0,0}(t), \dots, \vec{\xi}_{2,2}(t), \hat\phi(t)$}                      \\ \cmidrule(l){2-3} 
		\multicolumn{1}{l}{}                                   & Below-rated wind speed & Above-rated wind speed \\ \midrule
		Maximum polynomial degree                              & 4                     & 4                      \\
		Interaction order                                      & 1                     & 1                      \\
		Auto-regressive lags                                   & $\{1\}$           & $\{1\}$            \\
		Exogenous input lags                                   & $\{1,2\}$               & $\{1,2\}$                \\
		Number of coefficients                                 & 84                    & 84                      \\
		Number of training samples                             & $\mathcal{O}(10^6)$                   & $\mathcal{O}(10^6)$                     \\ \bottomrule
	\end{tabular}
	\label{tab:rotor_speed_mNARX_config}
\end{table}

Finally, to predict the generated power $P$, we again create a new exogenous input manifold $\vec{\zeta}_P(t) = \{\vec{\xi}_P(t), \hat\phi(t), \hat\omega(t)\}$ which includes a subset of the spectral coefficients (Eq.~\eqref{eq:discrete_cosine_transform}), the predicted blade pitch and the predicted rotor speed.
The NARX model based on this input then reads
\begin{equation}
	\hat P(t) = \hat{\mathcal{M}}_P^\text{}(\vec{\zeta}_P(\mathcal{T}\le t))
\end{equation}
and its specific structure is reported in Tab.~\ref{tab:genpwr_mNARX_config}.
\begin{table}[htb]
	\centering
	\caption{Wind turbine simulation -- Configuration of the NARX surrogate for the prediction of the generator power $P(t)$}
	\begin{tabular}{@{}lcc@{}}
		\toprule
		\multicolumn{1}{l}{\multirow{2}{*}{Exogenous inputs:}} & \multicolumn{2}{c}{$\vec{\xi}_{0,0}(t), \dots, \vec{\xi}_{1,1}(t), \hat\phi(t), \hat\omega(t)$}                      \\ \cmidrule(l){2-3} 
		\multicolumn{1}{l}{}                                   & Below-rated wind speed & Above-rated wind speed \\ \midrule
		Maximum polynomial degree                              & 7                     & 3                      \\
		Interaction order                                      & 1                     & 3                      \\
		Auto-regressive lags                                   & $\{1,30\}$           & $\{1,30\}$            \\
		Exogenous input lags                                   & $\{1,30\}$               & $\{1\}$                \\
		Number of coefficients                                 & 98                    & 164                      \\
		Number of training samples                             & $\mathcal{O}(10^6)$                   & $\mathcal{O}(10^5)$                     \\ \bottomrule
	\end{tabular}
	\label{tab:genpwr_mNARX_config}
\end{table}

Our second main QoI, the flapwise blade root moment $M^\text{Bld}$ strongly depends on the blade pitch and the azimuth $\alpha(t)$ of the blades. 
When $\alpha$ is zero (the first blade pointing upwards) the blade is axially under compression because of gravity and typically experiences higher winds at this higher altitude. With $\alpha=180$° the blade points down, is under tension and exposed to lower wind speeds. 
Consequently, $M^\text{Bld}$ is periodic when the rotor speed is constant and quasi-periodic in practice when the rotor speed varies. $M^\text{Bld}$ shows even higher periodicity due to \textit{tower shadowing} effect when any of the three blades pass the turbine tower.
To account for these phenomena we integrate the predicted rotor speed in time to obtain a prediction for the rotor azimuth $\hat\alpha(t) = \int_0^t \hat\omega(t) \text{d}t$, and construct the higher harmonics $\hat{\vec{z}}(t) = \{ \cos{(k\hat\alpha(t))}, \sin{(k\hat\alpha(t))} \}$, ${k=\{1,\dots,4\}}$ from it.
These harmonics, in conjunction with a large set of spectral coefficients and the blade pitch prediction, build a new manifold $\vec{\zeta}_{M^\text{Bld}} = \{ \vec{\xi}_{M^\text{Bld}}(\mathcal{T} \le t), \hat{\phi}(\mathcal{T} \le t), \hat{\vec{z}}(\mathcal{T} \le t) \}$ that allows us to model the complex evolution of $M^\text{Bld}$:
\begin{equation}
	\hat M^\text{Bld}(t) = \hat{\mathcal{M}}_{M^\text{Bld}}^\text{}(\vec{\zeta}_{M^\text{Bld}}(\mathcal{T} \le t)).
\end{equation}
As the blade moments depend on the wind speeds in its close proximity, a relatively large number of 25~spectral coefficients is used.
The full NARX configuration is given in Tab.~\ref{tab:blade_pitch_mNARX_config}
\begin{table}[htb]
	\centering
	\caption{Wind turbine simulation -- Configuration of the NARX surrogate for the prediction of the flapwise blade root moment $M^\text{Bld}$}
	\begin{tabular}{@{}lcc@{}}
		\toprule
		\multicolumn{1}{l}{\multirow{2}{*}{Exogenous inputs:}} & \multicolumn{2}{c}{$\vec{\xi}_{0,0}(t), \dots, \vec{\xi}_{4,4}(t), \hat\phi(t), \cos{(k\hat\alpha(t))}_{k=\{1, \dots, 4\}}, \sin{(k\hat \alpha(t))}_{k=\{1, \dots, 4\}}$}                      \\ \cmidrule(l){2-3} 
		\multicolumn{1}{l}{}                                   & Below-rated wind speed & Above-rated wind speed \\ \midrule
		Maximum polynomial degree                              & 7                     & 3                      \\
		Interaction order                                      & 1                     & 3                      \\
		Auto-regressive lags                                   & $\{1,30\}$           & $\{1,30\}$            \\
		Exogenous input lags                                   & $\{1,30\}$               & $\{1\}$                \\
		Number of coefficients                                 & 740                    & 702                      \\
		Number of training samples                             & $\mathcal{O}(10^5)$      &            $\mathcal{O}(10^5)$                     \\ \bottomrule
	\end{tabular}
	\label{tab:blade_moment_mNARX_config}
\end{table}

The results for the auxiliary quantities and main quantities of interest are presented in the following sections, namely Section~\ref{sec:blade_pitch_results}-\ref{sec:blade_moment_results}. 
Similar to the spring-mass system (Section~\ref{sec:spring_mass_system}), the autoregressive predictions are initialized with the true initial conditions for better validation as discussed in Section~\ref{sec:ar_modelling_prediction}. 
This is particularly important for the blade moment, as it oscillates and its phase depends on the rotor azimuth. 
Starting the prediction with a different initial state can lead to a different evolution of the blade moment. 
It is worth noting, however, that this is also the case for the ASE simulator, where different initial conditions, such as initial rotor speed and azimuth, will result in a different evolution of the system.

\subsubsection{Blade pitch}\label{sec:blade_pitch_results}
\begin{figure}[htb]
	\centering
	\begin{subfigure}[b]{0.49\textwidth}
		\centering
		\includegraphics[width=0.5\textwidth]{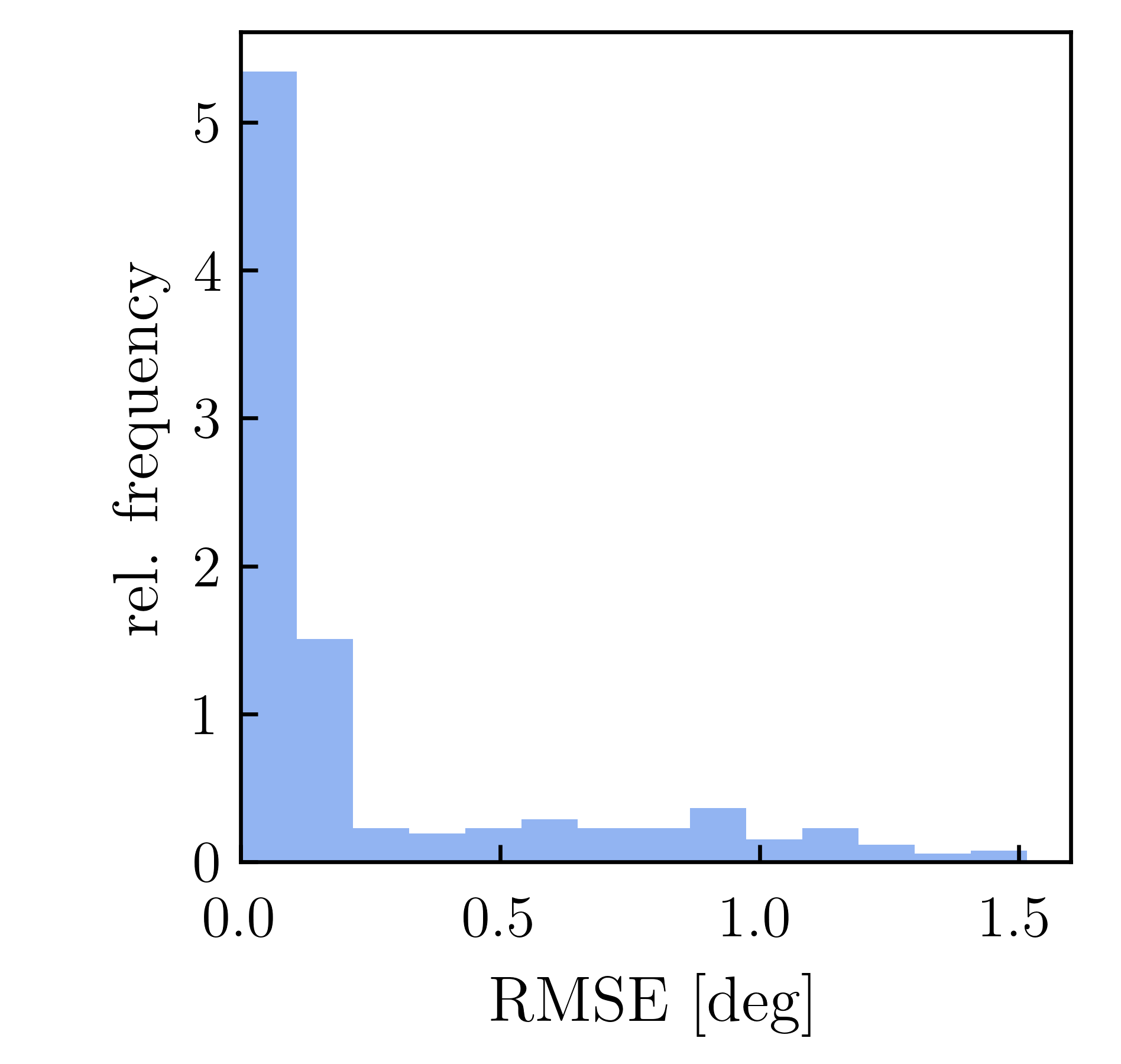}
	\end{subfigure}
	\begin{subfigure}[b]{0.49\textwidth}
		\centering
		\includegraphics[width=0.5\textwidth]{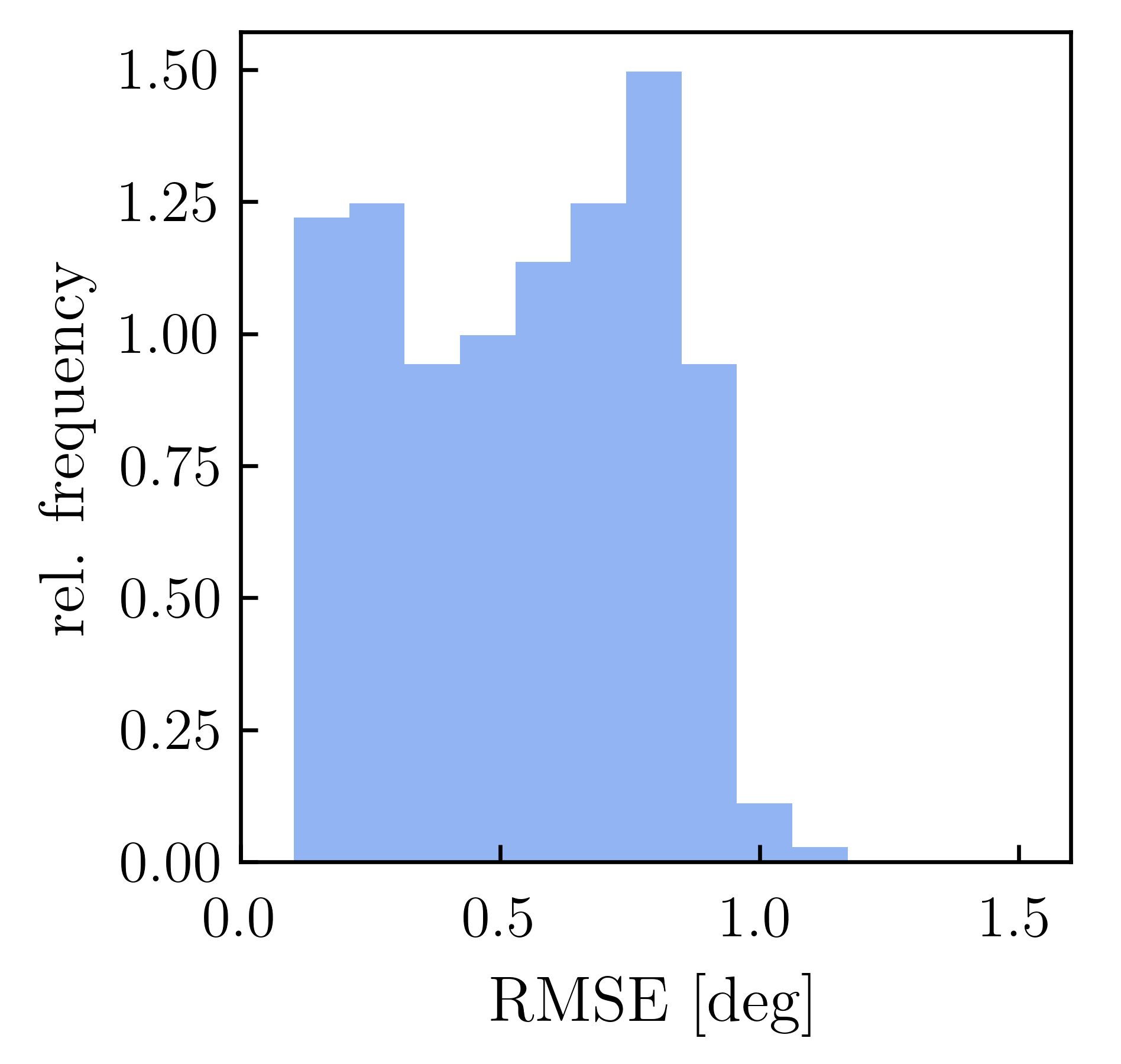}
	\end{subfigure}
	
	\begin{subfigure}[b]{0.49\textwidth}
		\centering
		\includegraphics[width=\textwidth]{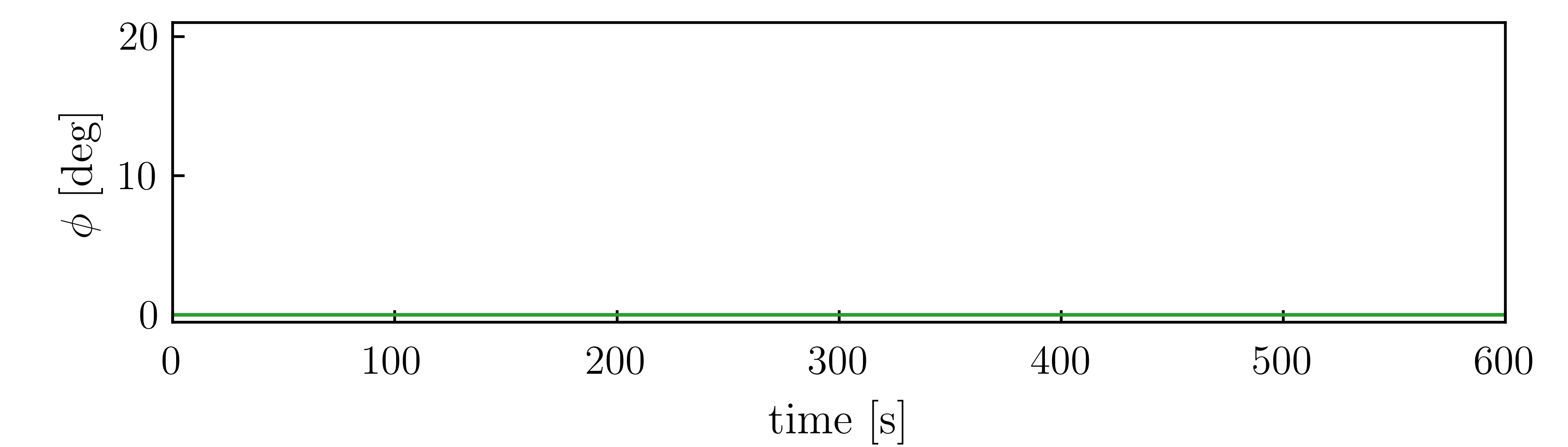}
	\end{subfigure}
	\begin{subfigure}[b]{0.49\textwidth}
		\centering
		\includegraphics[width=\textwidth]{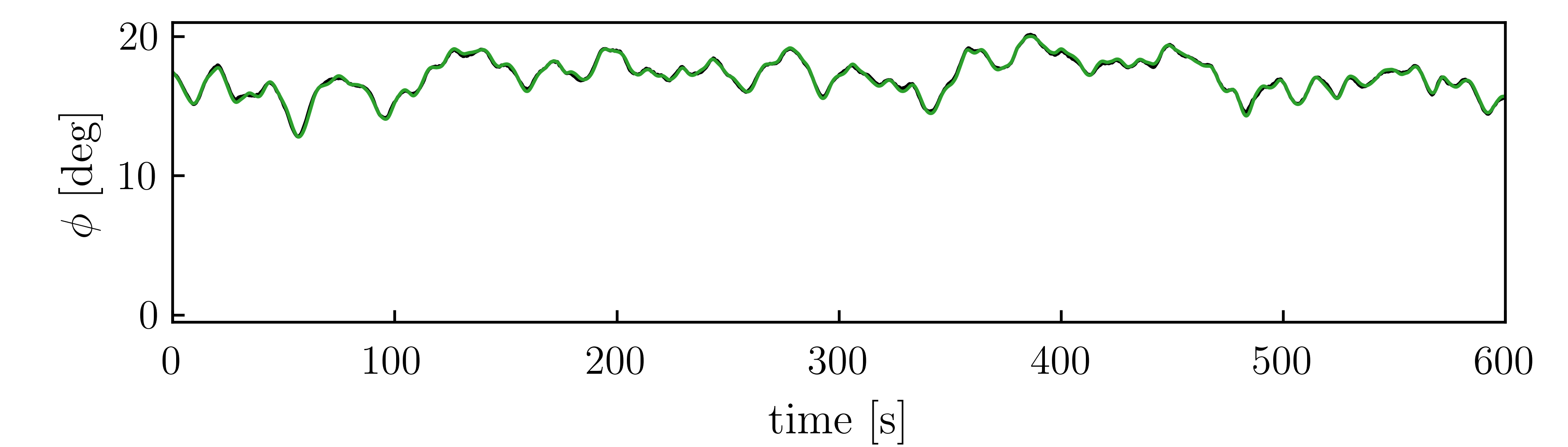}
	\end{subfigure}
	
	\begin{subfigure}[b]{0.49\textwidth}
		\centering
		\includegraphics[width=\textwidth]{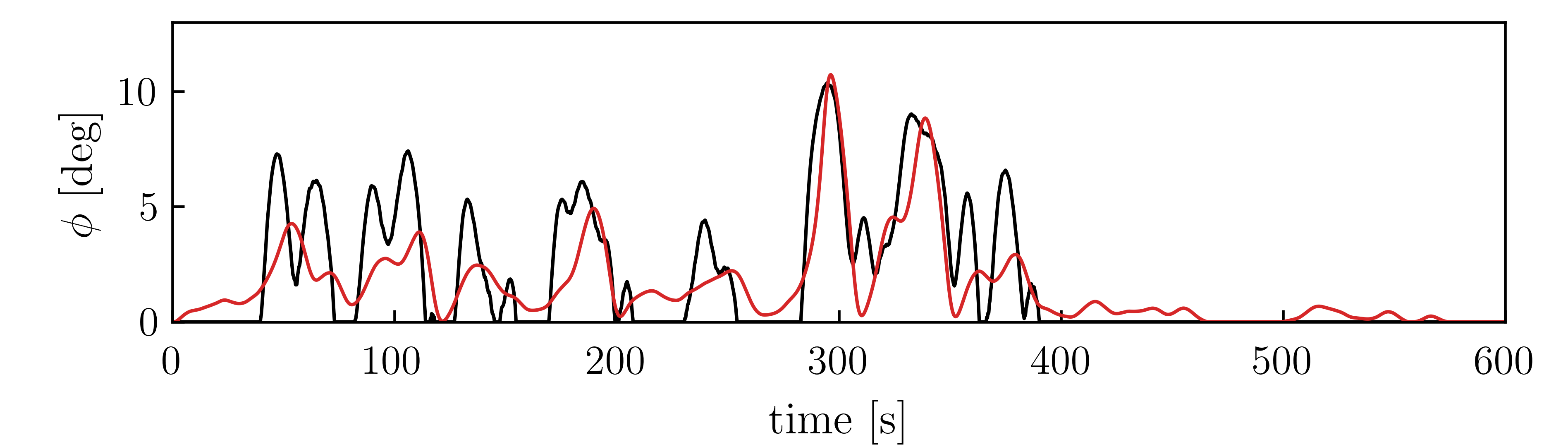}
		\caption{Below-rated wind speed regime}
	\end{subfigure}
	\begin{subfigure}[b]{0.49\textwidth}
		\centering
		\includegraphics[width=\textwidth]{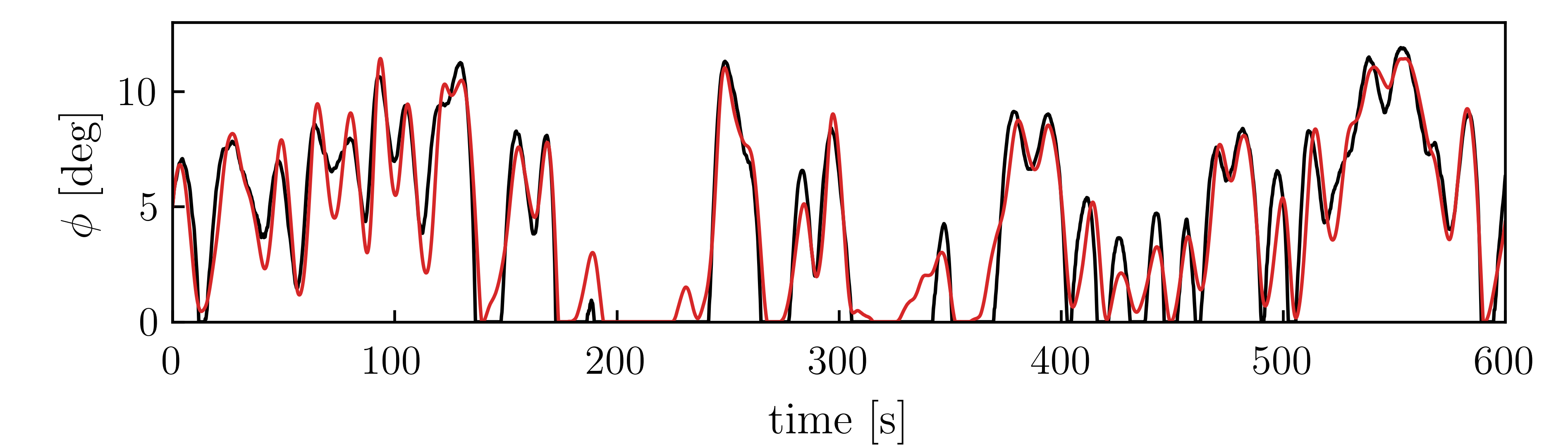}
		\caption{Above-rated wind speed regime}
	\end{subfigure}
	\caption{Wind turbine simulation -- Results for the \textbf{blade pitch} prediction in the below-rated (a) and above-rated wind speed regime (b).
		The top panel shows the root-mean-squared error (RMSE) of the blade pitch prediction in degrees on the validation dataset.
		The middle panel displays the true (black) and predicted (green) trace corresponding to the simulation with the lowest RMSE. 
		The bottom panel illustrates the traces corresponding to the simulation with the highest RMSE. The prediction is depicted in red and the true response in black.
	}
	\label{fig:blade_pitch_results}
\end{figure}

\newcommand{\phifig}[1]{Fig.~\ref{fig:blade_pitch_results}{#1}}
The performance of the NARX model on the first auxiliary quantity, namely the blade pitch $\phi(t)$, is shown in \phifig{}. \phifig{a} shows the results in the below-rated wind speed regime and \phifig{b} the above-rated wind speed regime. 

We report the root-mean-squared error (RMSE) in degrees in the top panels. 
In both regimes, the RMSE is low with a maximum of 1.5° and does not show any clear outliers. 
The high peak for the below-rated wind speed regime at zero RMSE can be explained by mostly having a zero pitch angle during the full 10~min simulation at low wind speeds. 
This region of the input space is also well captured by the surrogate. 

In the middle panels we show the traces with the lowest RMSE (best-case point of the validation set) for either regime. 
In the below-rated wind speed regime this is a simulation without any pitch controller action ($\phi=0$ over 600~s), and therefore modelled precisely. 
In the high wind speed regime, we see constantly high pitch angles between 10° and 20° which is replicated by the surrogate with very high accuracy. 
Similarly, we show the two samples with the highest RMSE (worst-case point of our validation set) in the bottom panels. 
It becomes clear that the surrogate for the low wind speed regime is not able to mimic the fast actions of the controller and returns a much smoother response. 
In the high wind speed regime where we see only a slightly more active controller, the surrogate exhibits a lower RMSE. 
This may be caused by the higher polynomial degree of the NARX model in this regime (see NARX configuration in Tab.~\ref{tab:model_configuration}) or because there is more data in this transition region from inactive to active pitch controller in the high wind speed training dataset.

\subsubsection{Rotor speed}
\begin{figure}[htb]
	\centering
	\begin{subfigure}[b]{0.49\textwidth}
		\centering
		\includegraphics[width=0.5\textwidth]{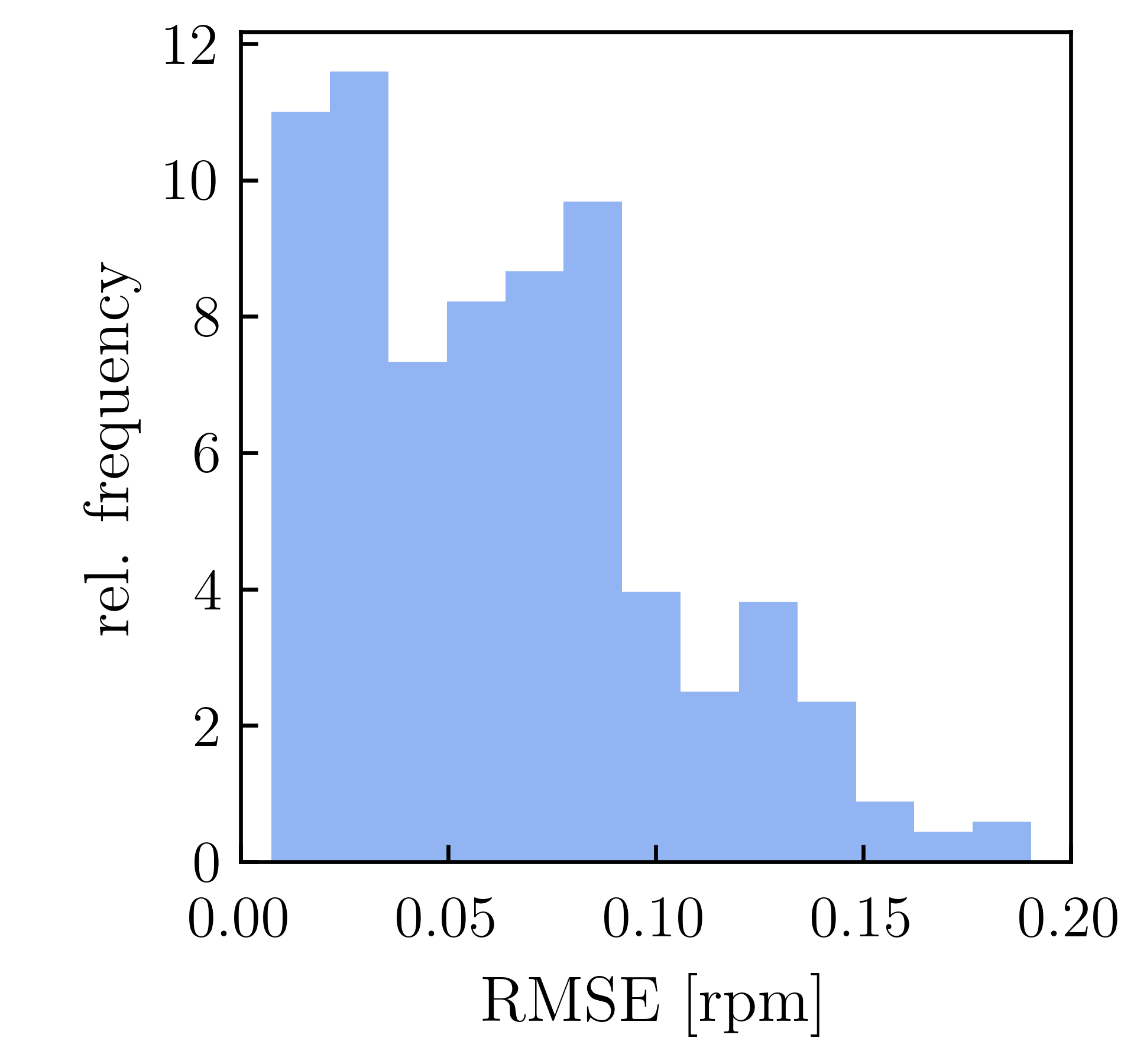}
	\end{subfigure}
	\begin{subfigure}[b]{0.49\textwidth}
		\centering
		\includegraphics[width=0.5\textwidth]{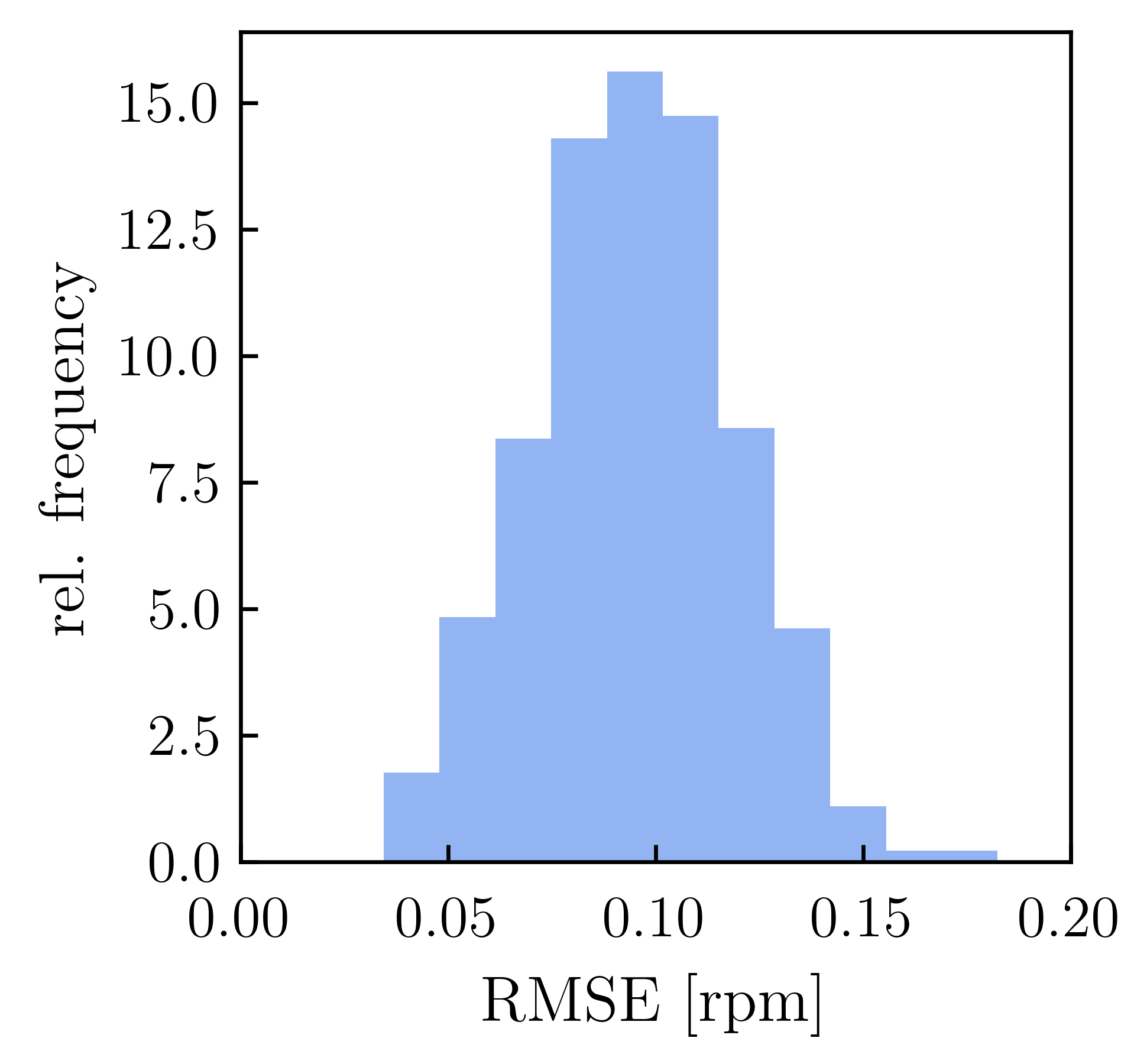}
	\end{subfigure}
	
	\begin{subfigure}[b]{0.49\textwidth}
		\centering
		\includegraphics[width=\textwidth]{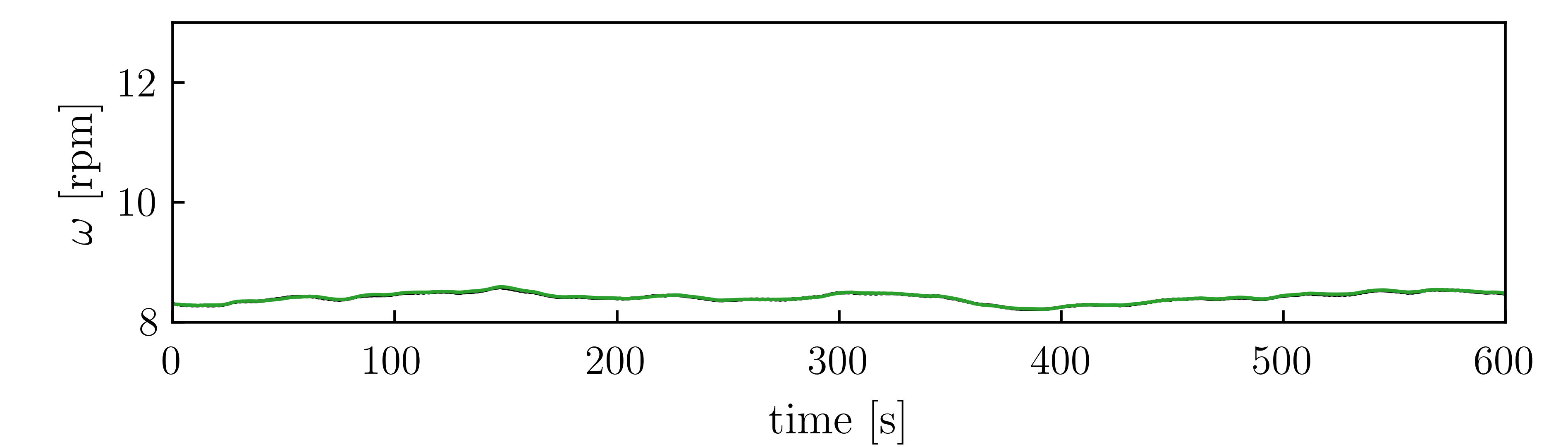}
	\end{subfigure}
	\begin{subfigure}[b]{0.49\textwidth}
		\centering
		\includegraphics[width=\textwidth]{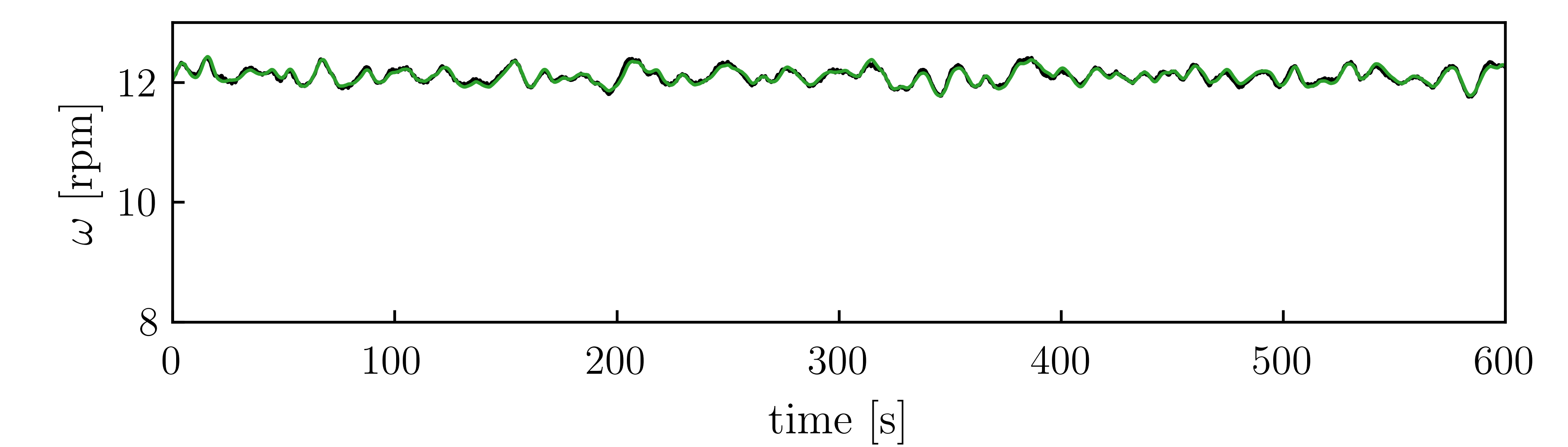}
	\end{subfigure}
	
	\begin{subfigure}[b]{0.49\textwidth}
		\centering
		\includegraphics[width=\textwidth]{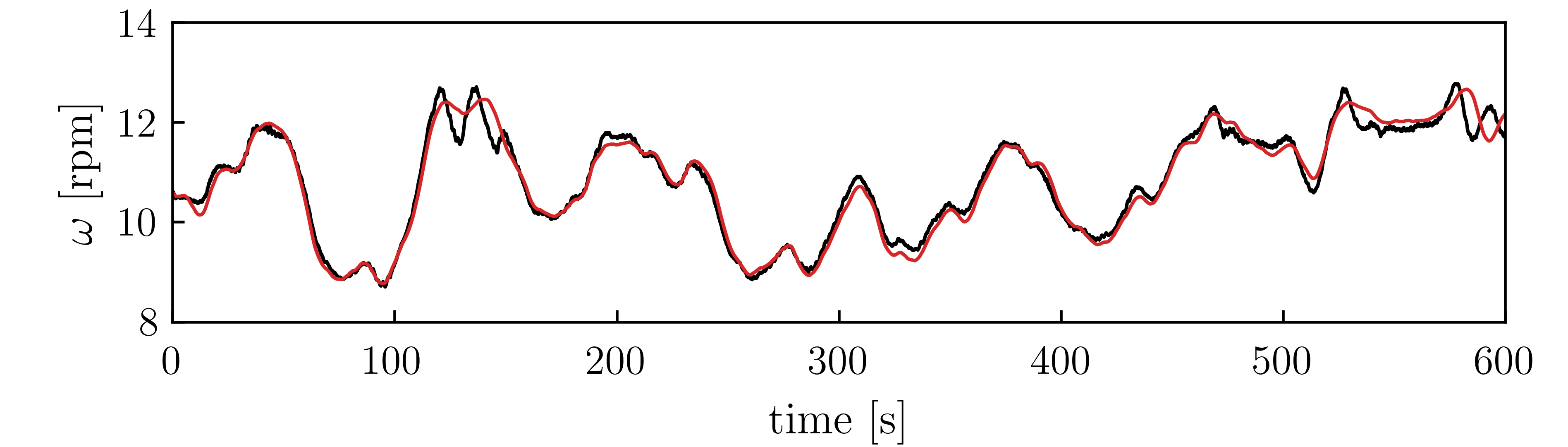}
		\caption{Below-rated wind speed regime}
	\end{subfigure}
	\begin{subfigure}[b]{0.49\textwidth}
		\centering
		\includegraphics[width=\textwidth]{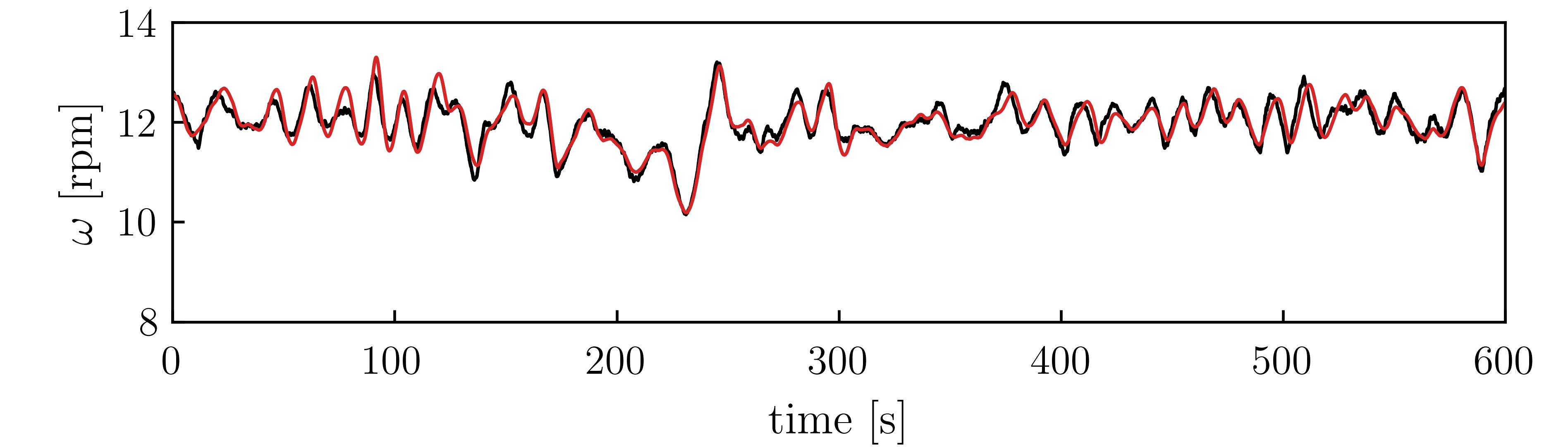}
		\caption{Above-rated wind speed regime}
	\end{subfigure}
	\caption{Wind turbine simulation -- Results for the \textbf{rotor speed} prediction in the below-rated (a) and above-rated wind speed regime (b).
		The top panel shows the root-mean-squared error (RMSE) of the rotor speed prediction in revolutions per minute on the validation dataset.
		The middle panel displays the true (black) and predicted (green) trace corresponding to the simulation with the lowest RMSE. 
		The bottom panel illustrates the traces corresponding to the simulation with the highest RMSE. The prediction is depicted in red and the true response in black.
	}
	\label{fig:rotor_speed_results}
\end{figure}

\newcommand{\omegafig}[1]{Fig.~\ref{fig:rotor_speed_results}{#1}}
Similarly to the blade pitch results we present the performance of mNARX on the second auxiliary quantity, i.e., the rotor speed $\omega(t)$. The top panels show a very low RMSE for the rotor speed prediction in both wind speed regimes. The relative frequency of the RMSE in the low wind speed regimes decreases with increasing RMSE, whereas for high wind speeds the RMSE is almost symmetrically distributed around an RMSE of 0.1~rpm. 
This very different distribution of the RMSE is related to the fact that at low wind speeds the rotor speed takes a wide range of values whereas it is kept at about the rated rotor speed of 12.1~rpm (see Tab.~\ref{tab:turbine_specs}) at high wind speeds (meaning about 1~\% accuracy).
Nevertheless, the middle panels, which show the simulations with the lowest RMSE (best-case), highlight the high accuracy of the surrogate. 

\clearpage

\subsubsection{Generator power}
\begin{figure}[htb]
	\centering
	\begin{subfigure}[b]{0.49\textwidth}
		\centering
		\includegraphics[width=\textwidth]{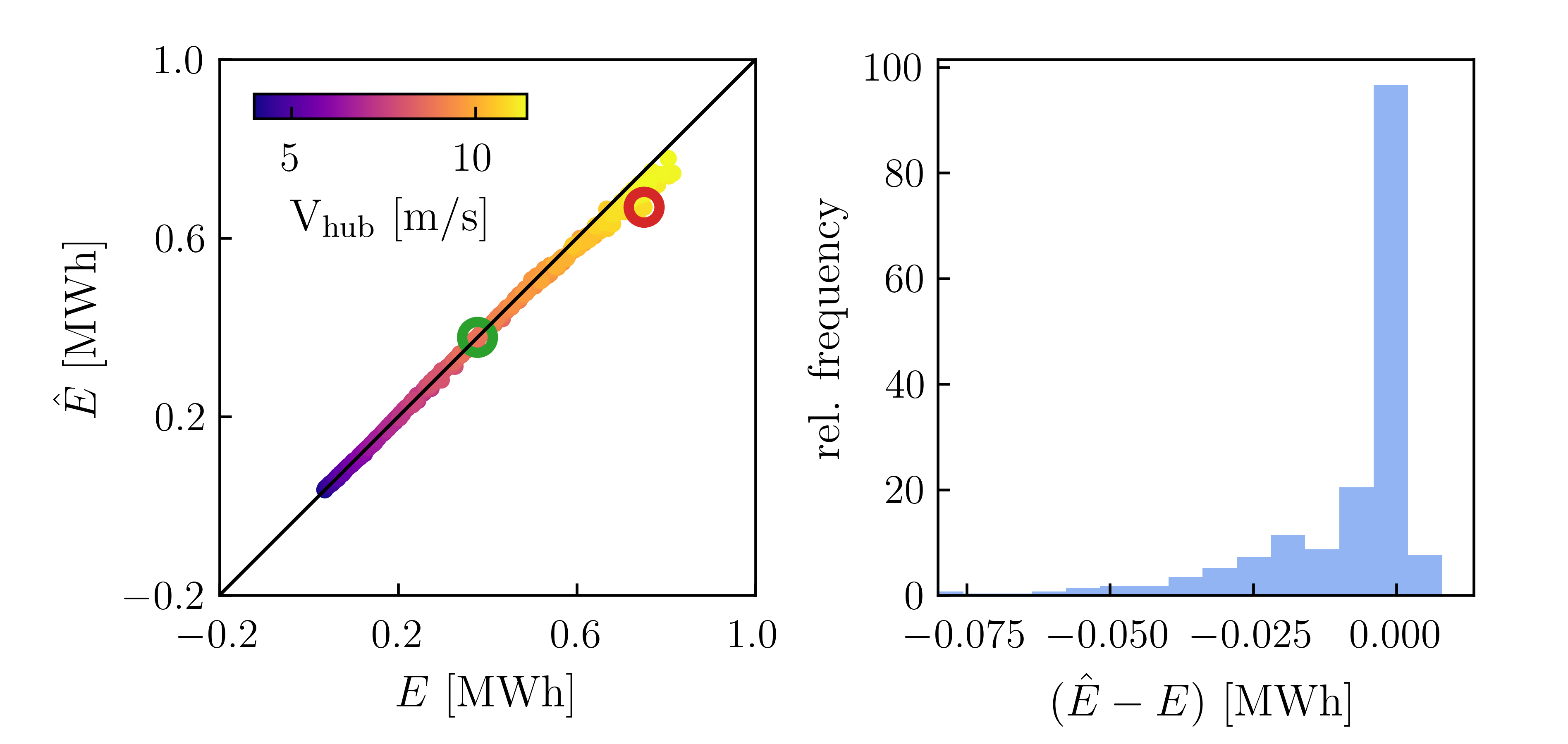}
	\end{subfigure}
	\begin{subfigure}[b]{0.49\textwidth}
		\centering
		\includegraphics[width=\textwidth]{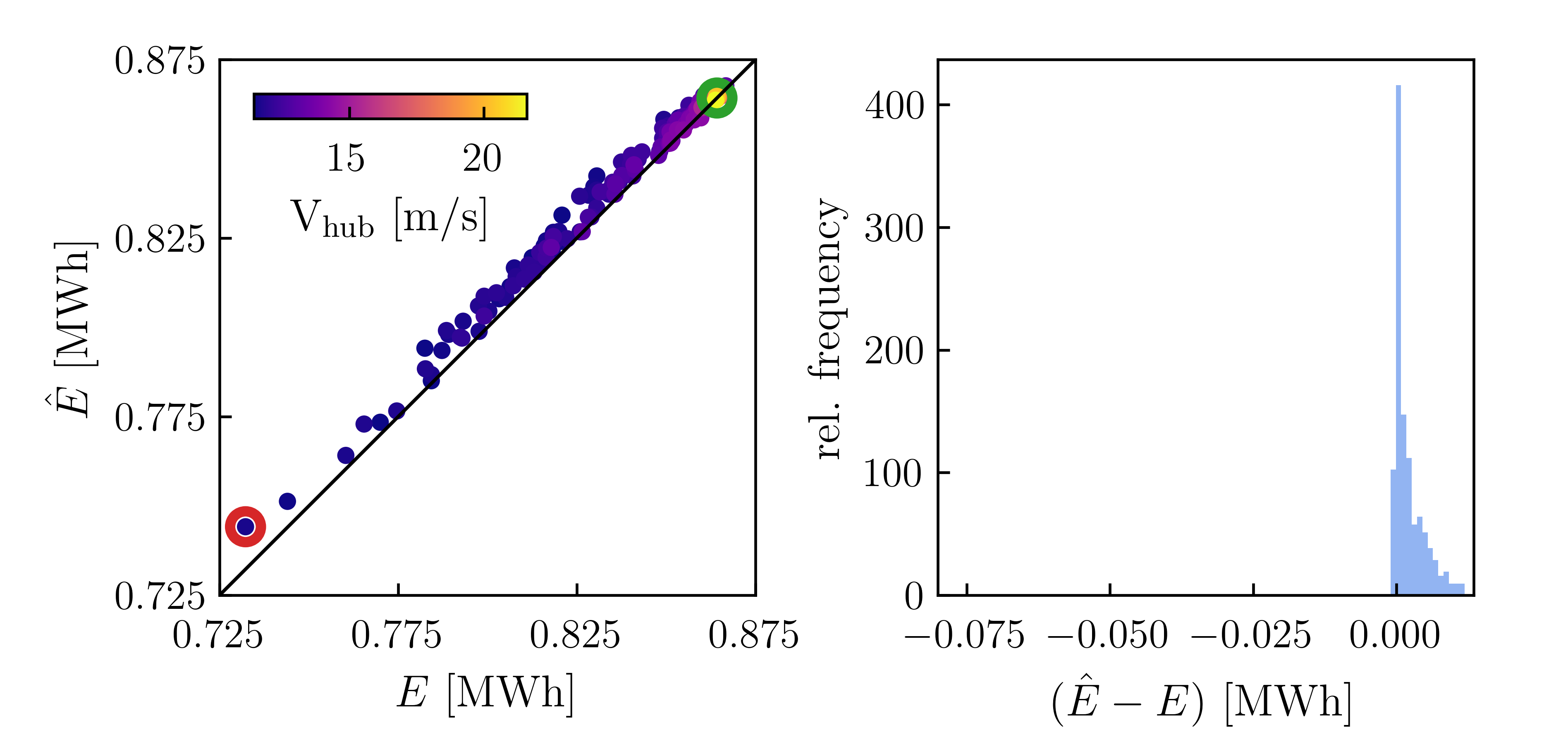}
	\end{subfigure}
	
	\begin{subfigure}[b]{0.49\textwidth}
		\centering
		\includegraphics[width=\textwidth]{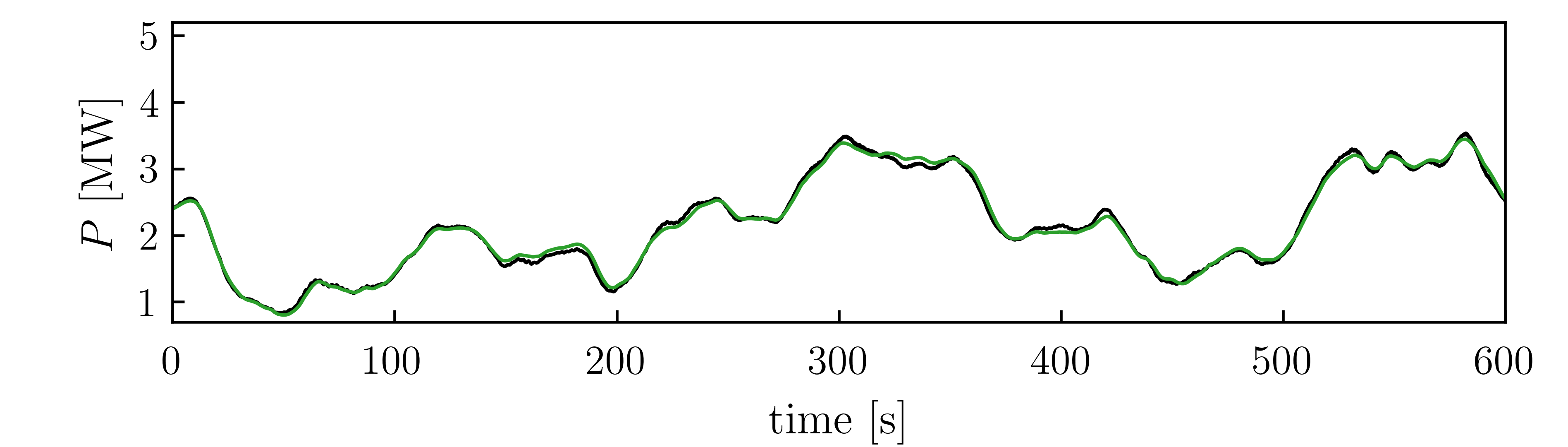}
	\end{subfigure}
	\begin{subfigure}[b]{0.49\textwidth}
		\centering
		\includegraphics[width=\textwidth]{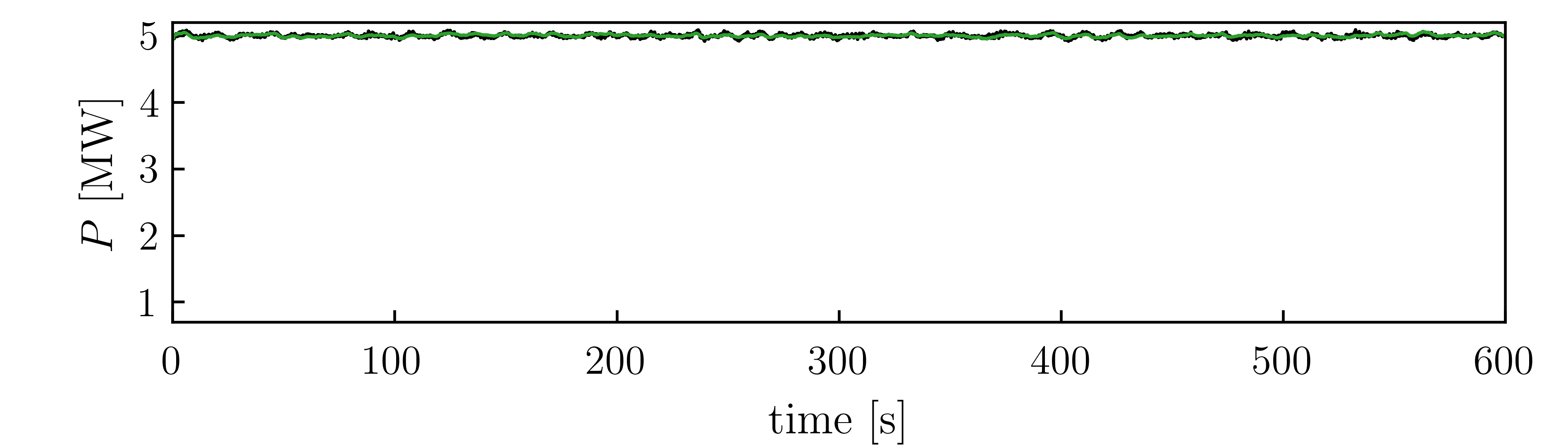}
	\end{subfigure}
	
	\begin{subfigure}[b]{0.49\textwidth}
		\centering
		\includegraphics[width=\textwidth]{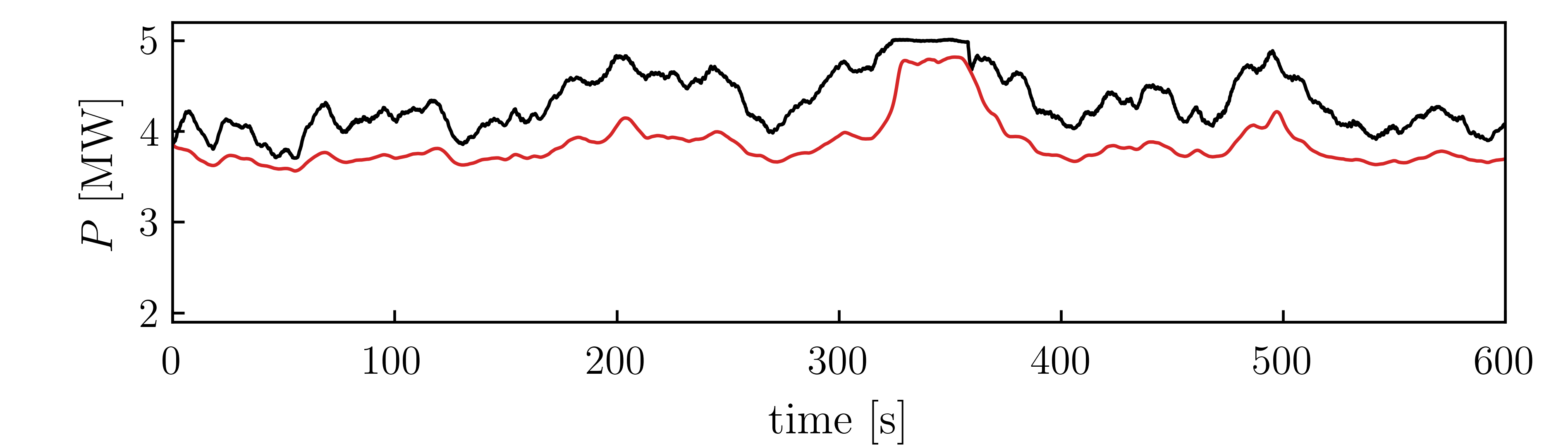}
		\caption{Below-rated wind speed regime}
	\end{subfigure}
	\begin{subfigure}[b]{0.49\textwidth}
		\centering
		\includegraphics[width=\textwidth]{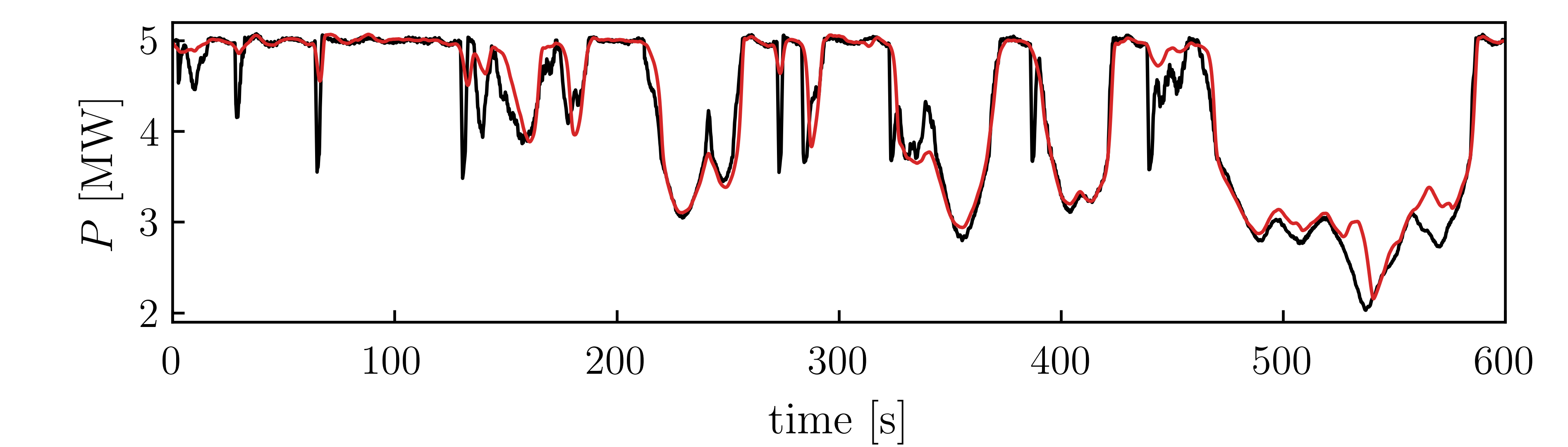}
		\caption{Above-rated wind speed regime}
	\end{subfigure}
	\caption{Wind turbine simulation -- Results for the \textbf{generator power} prediction in the below-rated (a) and above-rated wind speed regime (b).
		The left top panel shows the produced energy of the true output $E$ and of the prediction $\hat E$ in MWh for each simulation in the validation dataset. The color of each scatter represents the reference wind speed $V_\mathrm{hub}$ of that simulation.
		The right top panel shows the histogram of the difference between predicted and true produced energy.
		The middle panel displays the true and predicted trace of the power output corresponding to the simulation with the lowest difference $\hat E - E$. The true response is shown in black and the prediction in green. 
		The bottom panel illustrates the traces corresponding to the simulation with the highest error. The prediction is shown in red and the true values in black.
	}
	\label{fig:generator_power_results}
\end{figure}

\newcommand{\genpwrfig}[1]{Fig.~\ref{fig:generator_power_results}{#1}}
The results for the generator power $P$ are given in \genpwrfig{}.
To provide a general overview of the results we plot the produced energy $E$ during the 10~min simulation of the true power production and the predicted one for the above- (\genpwrfig{a}) and below-rated wind speed regime (\genpwrfig{b}) in the top left panels. 
The heatmap represents the reference wind speed $V_\text{hub}$ of each simulation. 
In the low wind speed regime, the energy production increases with increasing $V_\text{hub}$. In contrast, in the high wind speed regime in most simulations the energy production reaches its saturation level of 0.83~MWh, which corresponds to the turbine operating for 10~min at its rated power of 5~MW. 
In the low wind speed regime, the surrogate tends to underpredict the power output, as can be seen clearly in the top right panel in \genpwrfig{b}, where we plot the difference between the true and predicted energy production ($\hat E - E$). 
In the high wind speed regime, there is a slight overprediction of the energy production as shown in the top right panel in \genpwrfig{b}. On average the energy production is overpredicted by 0.0021~MWh which is 0.25~\% of the saturation level.

The two middle panels display the simulations with the lowest discrepancy in the $E$. The surrogates predict the true power output well in both subplots. 
The bottom panel in \genpwrfig{a} confirms the findings from the discrepancy plot that for high errors the surrogate consistently underestimates the power output in the below-rated wind speed regime.
This mostly happens at wind speeds close to the classification boundary which also corresponds to the region in which the turbine reaches its rated power. This supports the indications from Section~\ref{sec:blade_pitch_results} that there is a lack of training samples in this input space region.
The bottom panel in \genpwrfig{b} which shows the worst prediction in the above-rated wind speed validation dataset shows still good agreement with the true output even when the power output drops well below rated power. 
The fact that the simulations in the two bottom panels have a similar $V_\text{hub}$ again supports the assumption that the below-rated wind speed data set is lacking enough samples close to or at rated power. 

\subsubsection{Flapwise blade root moment results}\label{sec:blade_moment_results}
\begin{figure}[htb]
	\centering
	\begin{subfigure}[b]{0.49\textwidth}
		\centering
		\includegraphics[width=\textwidth]{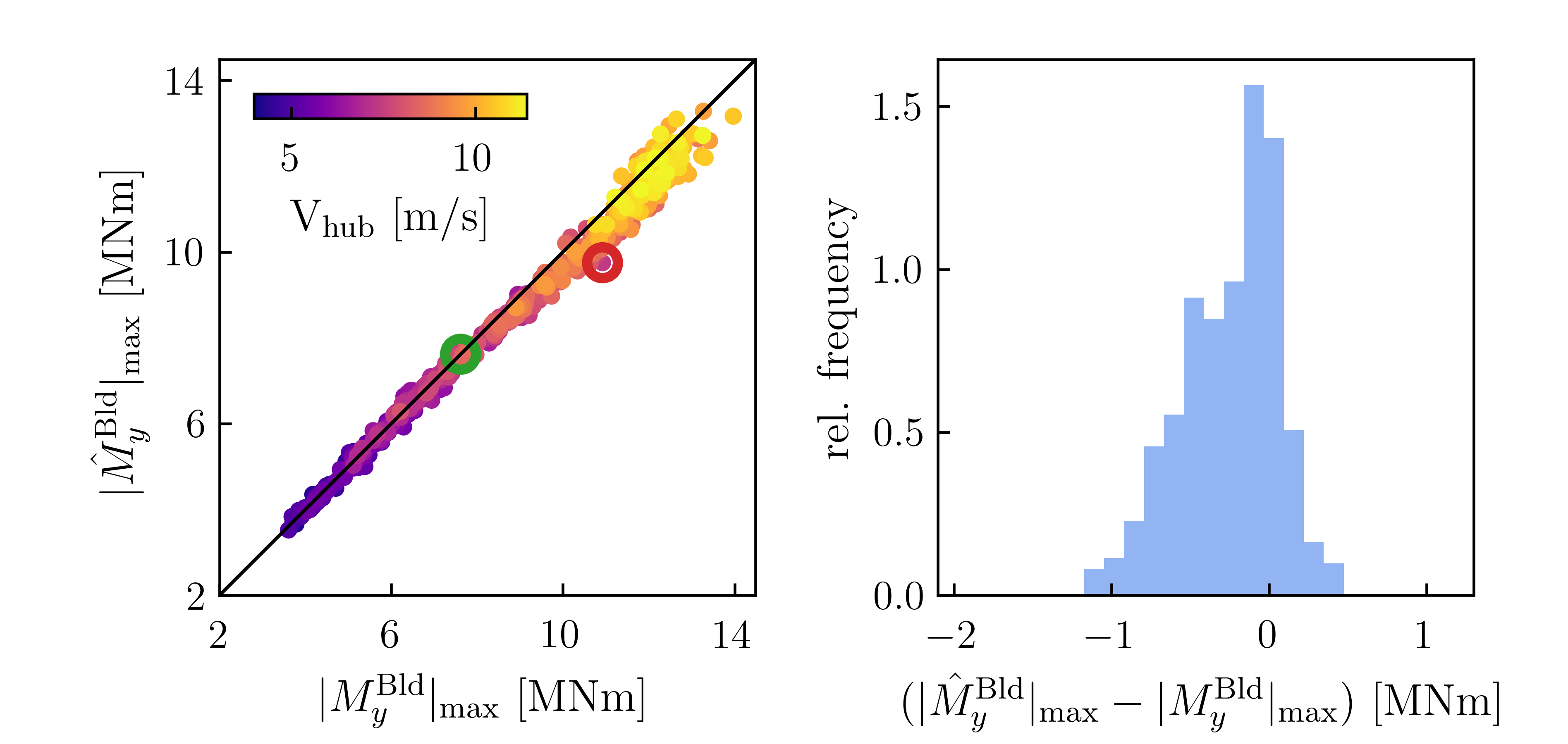}
	\end{subfigure}
	\begin{subfigure}[b]{0.49\textwidth}
		\centering
		\includegraphics[width=\textwidth]{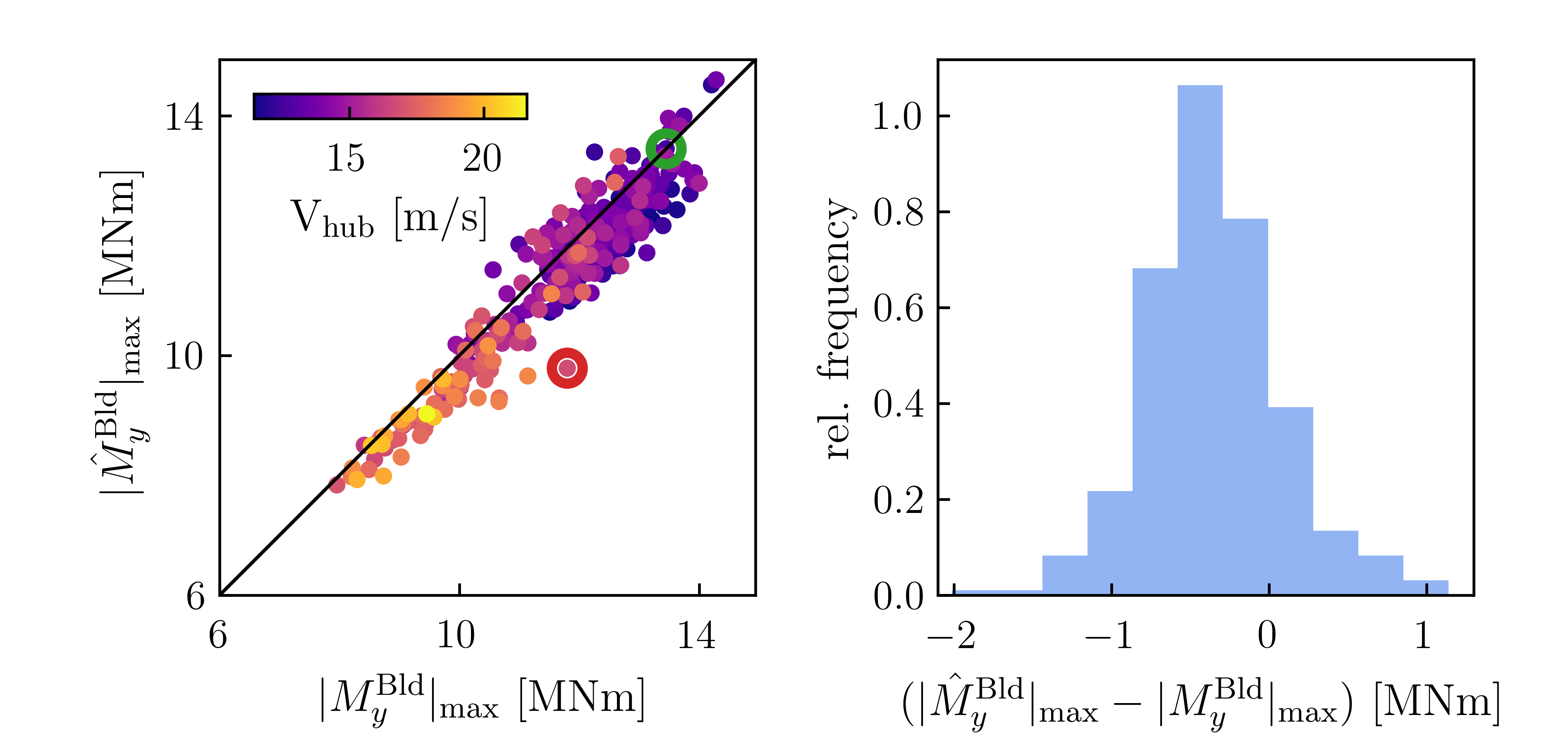}
	\end{subfigure}
	
	\begin{subfigure}[b]{0.49\textwidth}
		\centering
		\includegraphics[width=\textwidth]{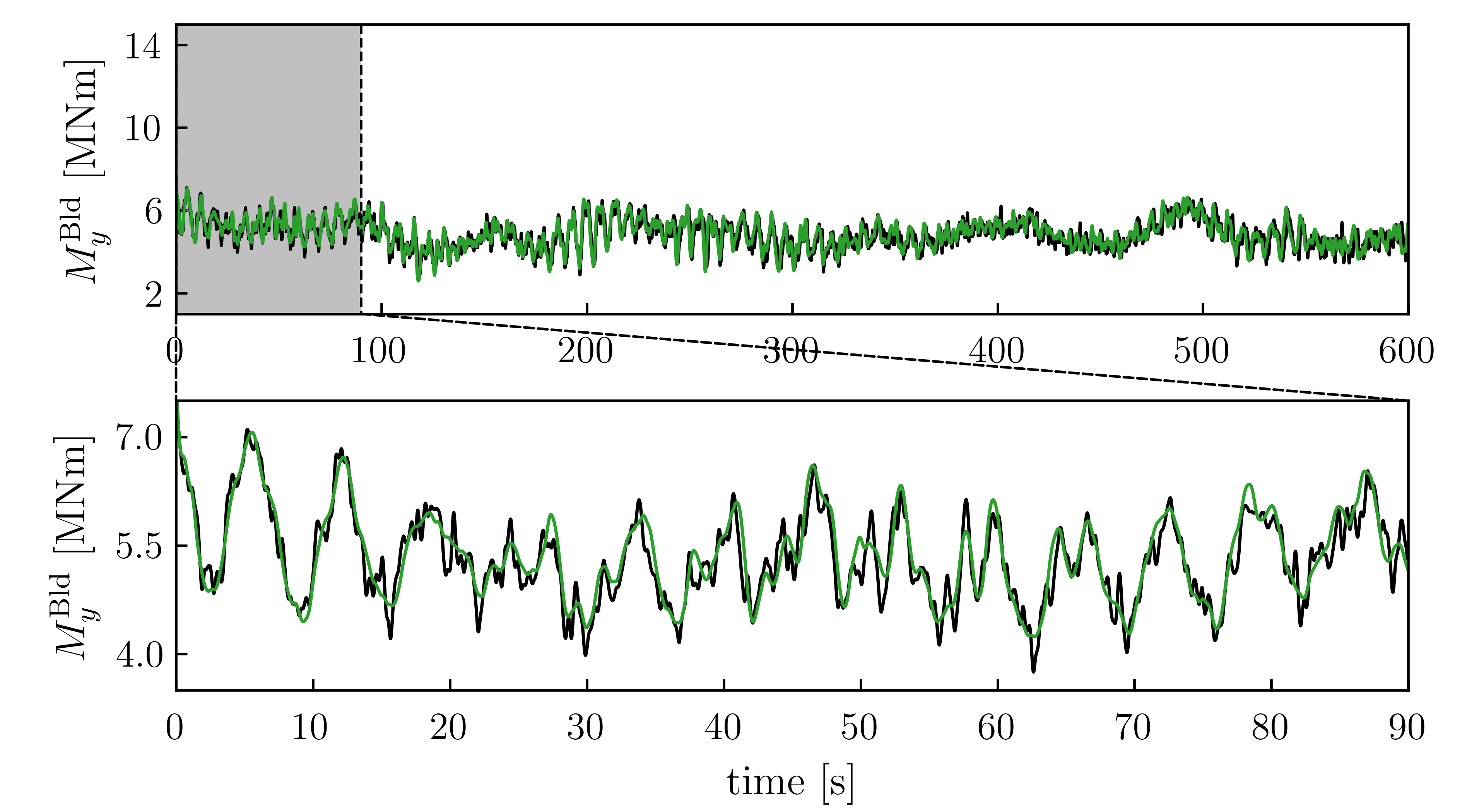}
	\end{subfigure}
	\begin{subfigure}[b]{0.49\textwidth}
		\centering
		\includegraphics[width=\textwidth]{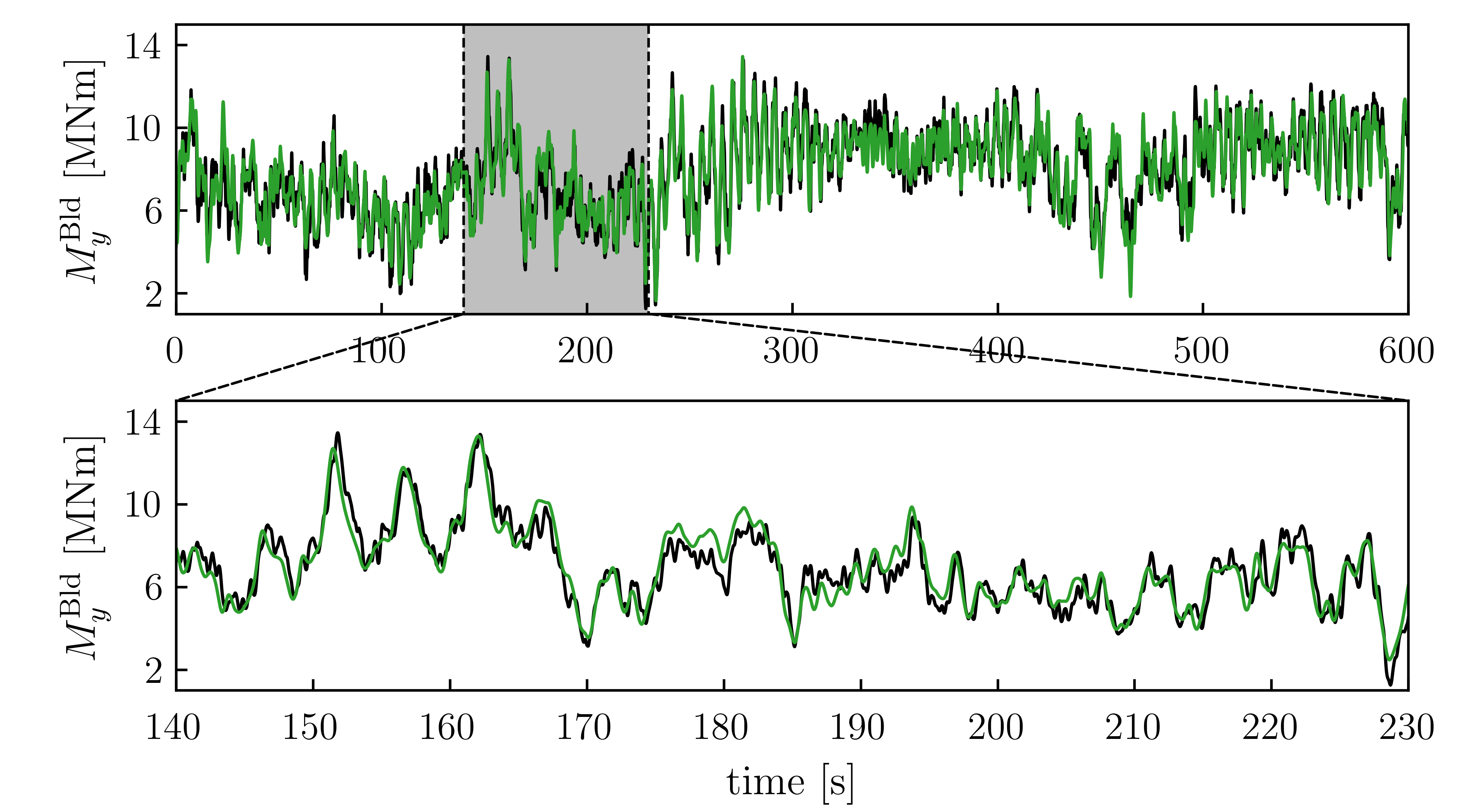}
	\end{subfigure}
	
	\begin{subfigure}[b]{0.49\textwidth}
		\centering
		\includegraphics[width=\textwidth]{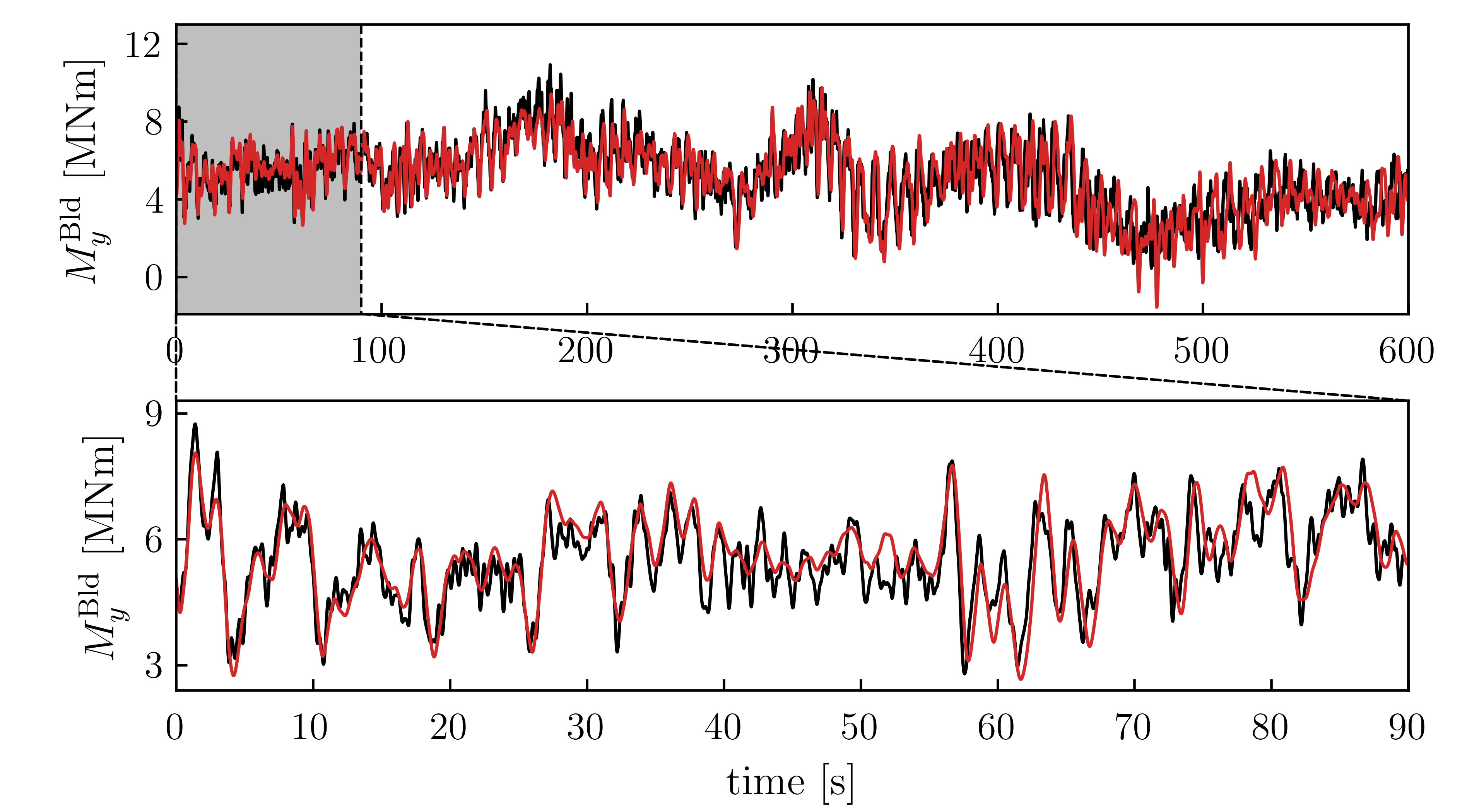}
		\caption{Below-rated wind speed regime}
	\end{subfigure}
	\begin{subfigure}[b]{0.49\textwidth}
		\centering
		\includegraphics[width=\textwidth]{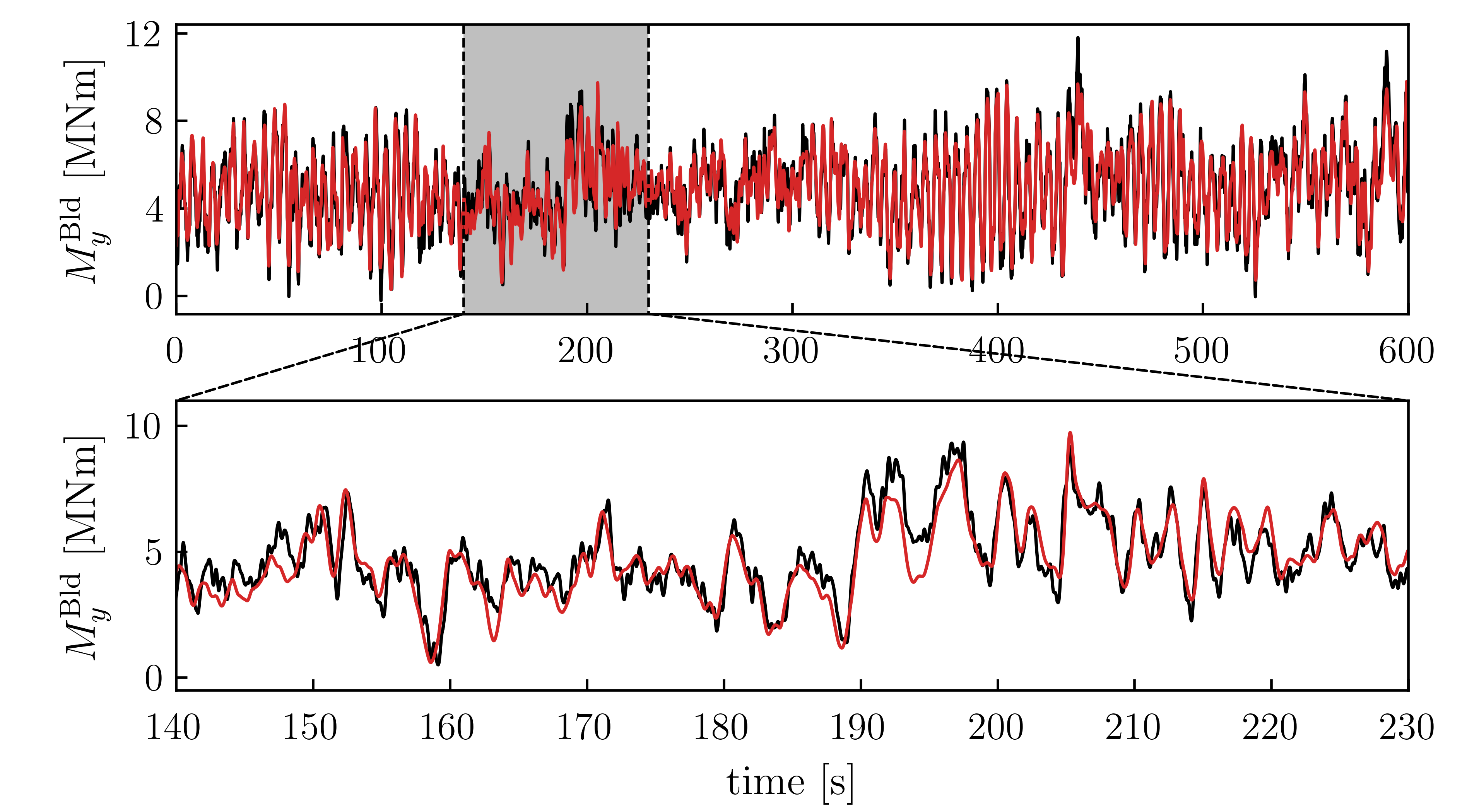}
		\caption{Above-rated wind speed regime}
	\end{subfigure}
	\caption{Wind turbine simulation -- Results for the \textbf{flapwise blade root moment} $M^\text{Bld}$ prediction in the below-rated (a) and above-rated wind speed regime (b).
		The left top panel shows the absolute peak moment $|M^\text{Bld}|_\mathrm{max}$ of the true response and of the prediction in MNm for the validation dataset. The color shows the reference wind speed $V_\mathrm{hub}$.
		The right top panel shows the histogram of the difference between predicted and true peak moment.
		The middle panel displays the true and predicted trace of the moment corresponding to the simulation with the lowest error in the maximum moment. The true response is shown in black and the prediction in green. 
		The bottom panel illustrates the traces corresponding to the simulation with the highest error. The prediction is shown in red and the true values in black.
	}
	\label{fig:blade_moment_power_results}
\end{figure}

\newcommand{\mybldfig}[1]{Fig.~\ref{fig:blade_moment_power_results}{#1}}
The results for the flapwise blade root moment $M^\text{Bld}$ in the below-rated wind speed regime are displayed in \mybldfig{a} and we present the result for the above-rated wind speed regime in \mybldfig{b}.
We quantify the accuracy of the surrogate in terms of the absolute peak moment $|M^\text{Bld}|_\text{max}$, which is one of the main quantities of interest in the wind turbine design process. 
Note that the peak value is extracted from the \emph{surrogated} time series as $|\hat M^\text{Bld}|_\text{max}$ (that is, we do not surrogate this scalar quantity directly, as it is common in the wind energy literature).

In the top left panels, we compare the peak moment of the simulator output $|M^\text{Bld}|_\text{max}$ and prediction of the surrogate $|\hat M^\text{Bld}|_\text{max}$. 
The heatmap represents the reference wind speed $V_\text{hub}$ of the simulation.
For the low wind speed regime we see good agreement between true and predicted peak moments.
For the high wind speed regime the accuracy is reduced, especially for wind speeds at about 17~m/s and above. 
These qualitative results are confirmed by the histograms in the top right panels which show the prediction discrepancy $|\hat M^\text{Bld}|_\text{max} - |M^\text{Bld}|_\text{max}$. 
The error on the below-rated wind speed validation dataset is about twice lower than on the above-rated wind speed validation dataset.

In the two middle panels the traces with the lowest error (best-case of the validation set) in the peak moment are displayed. The 90~s long sections marked in grey cover the true peak moment and are shown in more detail below the full 600~s traces. 
In the low wind speed regime the predicted peak value matches the true one almost exactly because the highest moment occurs at the beginning of the prediction. Note that as explained in Sec~\ref{sec:mNARX_structure} we initialize the prediction with the true values for a better comparison and therefore no error built up yet. 
In the high wind speed regime the true peak value appears later in the simulation after around 150~s while the predicted peak value occurs about 10~s after the true one. 
However, because the two peaks in the data are of similar magnitude, the error remains small.

In the two bottom panels, the traces with the highest error (worst-case of the validation set) are displayed.
From the trace of the low wind speed regime it becomes clear that an error in the rotor speed can propagate to the prediction of the blade moment as stated in Sec~\ref{sec:mNARX_structure}.
There is a clear mismatch in the phase of the true and predicted moment which is caused by a wrong azimuth, which is derived from the predicted rotor speed (see Section~\ref{sec:mNARX_structure}).
A different pattern can be observed in the above-rated wind speed regime where the phases do match even after more than 400~s but the sharp peak in the response is still underpredicted.
Nevertheless, the two traces with the highest error still show stable predictions over the full 600~s and visually agree well with the true response.

\clearpage

\subsubsection{Discussion on numerical performance}\label{sec:numerical_performance}
Surrogate models are designed to be efficient alternatives to time-consuming computational models. They leverage the inherent regularity in the behavior of the computational model to produce comparable outputs at significantly reduced computational costs. 
Nevertheless, they are trained on a relatively small set of full model evaluations, the experimental design. 
Consequently, surrogates bring performance benefits when many evaluations of the full model are required for an analysis, and the costs of generating the experimental design are offset by the performance of the surrogate.

To show the performance benefits of mNARX over the aero-servo-elastic simulator, Tab.~\ref{tab:performance} displays the time required to obtain a certain number of evaluations of all quantities of interest (QoIs), namely: blade pitch, rotor speed, generator power and blade moment. 
The time required by the simulator to perform the evaluations is given in the second column, while the time required by the mNARX surrogate is given in the third. The speed ratio of the surrogate over the simulator is given in parentheses.
The last column lists the time required for the surrogate, considering training time, which includes the time required to create the experimental design and fit the surrogates for all QoIs in both wind speed regimes.
Note that this additional training time is performed only once, and the surrogate can then be applied to multiple analyses if used under conditions similar to the training conditions.

Tab.~\ref{tab:performance} shows that the mNARX surrogate evaluates a new input more than $400$ times faster than the simulator. 
Even accounting for the training of the surrogate, it provides considerable benefits for as few as $1,000$ evaluations. As the number of evaluations increases, the proportion of the initial cost of creating the experimental design and training the surrogate decreases, so the relative speed advantage increases.
Since the OpenFAST simulator is already a fast low-fidelity simulator, these numbers are conservative estimates. The speedup may be even more significant if a higher fidelity simulator is used or if an optimized experimental design strategy is developed to reduce the size of the experimental design.

\begin{table}[htb]
	\centering
	\caption{
		Time required to obtain various numbers of model evaluations. The second column lists the time required by OpenFAST. The third column lists the times required by the mNARX surrogate. The speed-up of using mNARX over the simulator is included in parenthesis. The last column additionally includes the time to generate the experimental design and train the mNARX surrogate. 
	}
	\begin{tabular}{@{}cccc@{}}
		\toprule
		\multicolumn{1}{l}{Number of evaluations} & OpenFAST {[}h{]} & mNARX {[}h{]} & mNARX incl. training {[}h{]} \\ \midrule
		$10^3$                                    & $76$               & $0.17$ ($436$)  & $15.4$ ($4.9$)                    \\
		$10^4$                                    & $760$              & $1.74$ ($436$)  & $17.0$ ($45$)                  \\
		$10^5$                                    & $7,600$            & $17.4$ ($436$) & $32.7$ ($233$)                  \\
		$10^6$                                    & $76,000$           & $174$ ($436$) &  $189$ ($401$)                \\ 
		\bottomrule
	\end{tabular}
	\label{tab:performance}
\end{table}

\section{Discussion and conclusions}
\label{sec:discussion_and_conclusion}
In this paper, we introduced mNARX, a novel surrogate modelling technique that enables the efficient and accurate emulation of the response of complex dynamical systems, even in the presence of e.g. active controllers.
To do so, mNARX sequentially builds a chain of Nonlinear AutoRegressive with eXogenous inputs (NARX) models. These models are trained on an incrementally built exogenous input manifold that can contain not only the raw exogenous input but also the prediction of the NARX models earlier in the chain and features derived from these predictions. 
Therefore, the number of available features, and thus the information content of the exogenous input, increases as the modelling chain becomes longer, allowing for the modelling of more intricate quantities of interest.

We demonstrate the effectiveness of mNARX on different case studies. 
In the first example, we emulate a coupled two-mass-two-spring system with a one-dimensional exogenous input, and show that mNARX is capable of emulating the response of both system components with high accuracy, despite being trained on an extremely small dataset. 
In the second case study, a full aero-servo-elastic wind turbine simulator, we demonstrate that mNARX can handle dynamical systems with high exogenous input dimensionality when combined with dimensionality reduction techniques. 
This case study further underscores the universal applicability of the mNARX algorithm, as it consistently delivers stable long-term predictions with relatively small error accumulation, even in scenarios with multiple auxiliary quantities such as the turbine control system and turbine state variables.

mNARX has the favourable property of requiring a small training dataset, since it capitalizes on several intermediate NARX models, each of which modelling a simpler sub-problem compared to the full problem.
Since each subproblem has relatively low complexity, we were able to use low-degree polynomial NARX models, and therefore keeping the computational cost of training and evaluating the full mNARX surrogate low.

Because this remarkable efficiency is achieved by decomposing the original problem into a set of relevant sub problems, mNARX works best when some prior knowledge of the system properties and underlying physics is available.
Especially in many applications in physics and engineering, extensive knowledge about the system is usually available, and additional knowledge can be obtained through data processing or measurements, making mNARX well-suited for these applications. 
In cases where this condition is not met, mNARX falls back to standard NARX modelling, i.e. gives at least as good results. 
It should be noted, however, that mNARX inherits some limitations of standard NARX modelling, e.g., it may fail in resonant systems when the system response is no longer governed by the exogenous excitation.

Our methodology is readily extendible in multiple directions. 
High-sampling rate time-series are often highly correlated, making it possible to subsample from the design matrix of the experimental design data (as explained in Section~\ref{sec:ar_modelling_fitting}). 
This not only speeds up the model calibration but also enhances accuracy in specific regions in the input or output domain. 
For instance, one may select more samples that are in proximity to extreme responses, if the primary focus is to capture extreme values (e.g. in view of reliability analysis).

To improve the performance of the mNARX surrogate in cases with high-dimensional exogenous inputs, nonlinear dimensionality reduction techniques can be employed.
Additionally, alternative NARX models, as an example neural-network based, may also enhance the predictive performance of the mNARX surrogate, albeit likely with an associated increase in computational cost and data consumption compared to polynomial models. 

To reduce the need for prior knowledge, ongoing research is focused on the \textit{automatic detection} and \textit{selection} of auxiliary quantities, as well as construction of the input manifold. 
A more data-driven setting can increase applicability in cases with limited knowledge of the system being modelled.

\section*{Acknowledgments}
This project is part of the HIghly advanced Probabilistic design and Enhanced Reliability methods
for high-value, cost-efficient offshore WIND (HIPERWIND) project and has received funding from the European Union's Horizon 2020 Research and Innovation Programme under Grant Agreement No. 101006689.

\bibliographystyle{chicago}
\bibliography{bibliography.bib}

\end{document}